\documentclass[aps,prd,letterpaper,showkeys,showpacs,twocolumn,nofootinbib,preprintnumbers,longbibliography,unsortedaddress]{revtex4-1}

\usepackage[top=2.8cm, bottom=2.8cm, left=2cm, right=2cm]{geometry}
\usepackage[utf8]{inputenc}  
\usepackage{amsmath,amssymb,lmodern,amsfonts} 
\usepackage{graphicx}
\usepackage{natbib}
\usepackage[caption=false]{subfig}
\usepackage{hyperref} 
\usepackage{color}
\usepackage[table]{xcolor}
\usepackage{natbib}
\usepackage{tabulary} 

\begin{document}

\preprint{PI/UAN-2020-675FT}

\title{Anisotropic Einstein Yang-Mills Higgs Dark Energy}

\author{J. Bayron Orjuela-Quintana}
\email{john.orjuela@correounivalle.edu.co}
\affiliation{Departamento  de  F\'isica,  Universidad  del Valle, \\ Ciudad  Universitaria Mel\'endez,  Santiago de Cali  760032,  Colombia}

\author{Miguel \'Alvarez}
\email{miguel.alvarez1@correo.uis.edu.co}
\affiliation{Escuela  de  F\'isica,  Universidad  Industrial  de  Santander, \\ Ciudad  Universitaria,  Bucaramanga  680002,  Colombia}

\author{C\'esar A. Valenzuela-Toledo}
\email{cesar.valenzuela@correounivalle.edu.co}
\affiliation{Departamento  de  F\'isica,  Universidad  del Valle, \\ Ciudad  Universitaria Mel\'endez,  Santiago de Cali  760032,  Colombia}

\author{Yeinzon Rodr\'iguez}
\email{yeinzon.rodriguez@uan.edu.co}
\affiliation{Centro de Investigaciones en Ciencias B\'asicas y Aplicadas, Universidad Antonio Nari\~no, \\ Cra 3 Este \# 47A-15, Bogot\'a D.C. 110231, Colombia}
\affiliation{Escuela  de  F\'isica,  Universidad  Industrial  de  Santander, \\ Ciudad  Universitaria,  Bucaramanga  680002,  Colombia}
\affiliation{Simons Associate at The Abdus Salam International Centre for Theoretical Physics, \\ Strada Costiera 11, I-34151, Trieste, Italy}


\begin{abstract}
In the context of the dark energy scenario, the Einstein Yang-Mills Higgs model in the SO(3) representation was studied for the first time by M. Rinaldi (see JCAP \textbf{1510}, 023 (2015)) in a homogeneous and isotropic spacetime. We revisit this model, finding in particular that the interaction between the Higgs field and the gauge fields generates contributions to the momentum density, anisotropic stress and pressures, thus making the model inconsistent with the assumed background. We instead consider a homogeneous but anisotropic Bianchi-I space-time background in this paper and analyze the corresponding dynamical behaviour of the system. We find that the only attractor point corresponds to an isotropic accelerated expansion dominated by the Higgs potential. However, the model predicts non-negligible anisotropic shear contributions nowadays, i.e. the current Universe can have hair although it will loose it in the future. We investigate the evolution of the equation of state for dark energy and highlight some possible consequences of its behaviour related to the process of large-scale structure formation. As a supplement, we propose the ``\emph{Higgs triad}" as a possibility to make the Einstein Yang-Mills Higgs model be consistent with a homogeneous and isotropic spacetime.
\end{abstract}



\maketitle

\section{Introduction} \label{Intro}

The accelerated expansion of the current Universe is an observational fact. First discovered by measurements on distant supernovae type Ia \cite{Riess:1998cb, Perlmutter:1998np}, and well stablished by other observations like the angular spectrum of the cosmic microwave background (CMB) temperature fluctuations \cite{Aghanim:2018eyx}, the distribution of large-scale structures \cite{Troxel:2017xyo,Abbott:2018xao}, and the baryon acoustic oscillations \cite{Sahni:2014ooa}, the origin of this accelerated expansion is yet to be known. A simple model agreeing with the aforementioned observations is $\Lambda$ Cold Dark Matter ($\Lambda$CDM), however, it is plagued by problems of both theoretical and observational nature \cite{Weinberg:1988cp,DiValentino:2019qzk,Verde:2019ivm,Handley:2019tkm}, opening the window to consider dynamical dark energy models as the possible explanation \cite{Guo:2018ans}. The alternatives to the cosmological constant $\Lambda$ have been split into two categories: the introduction of exotic forms of matter and modified gravity theories. While the latter has been recently under observational pressure \cite{Collett:2018gpf,Ezquiaga:2018btd,He:2018oai,Do:2019txf,Abbott:2018lct,Akiyama:2019cqa}, in the former little has been done in the search for a mechanism that drives the accelerated expansion while involving standard forms of matter minimally coupled to the Einstein gravity.

In an attempt to describe the dark energy mechanism using matter fields included in the Standard Model of Particle Physics \cite{Kane:1987gb}, the Higgs dark energy model was proposed in \cite{Rinaldi:2014yta}. In that model, a complex Higgs doublet charged under the SU(2) gauge group drives the accelerated expansion by rotating around the minimum of its potential, in the same way as in the ``spintessence" scenario \cite{Boyle:2001du}. This is possible by neglecting the interaction between the Higgs doublet and the gauge vector fields. Nonetheless, it was shown in \cite{Alvarez:2019ues} that if this interaction is properly considered, it yields to inconsistencies with the Friedmann-Lema\^{\i}tre-Robertson-Walker (FLRW) spacetime. These problems can be alleviated by gauge fixing the Higgs field, allowing a new dark energy mechanism and drastically changing the conclusions of the original work \cite{Alvarez:2019ues}.\footnote{Refs. \cite{Emoto:2002fb,Hosotani:2002nq} had already shown that appropriately fixing the gauge for the Higgs field makes the system compatible with a FLRW background although the incompatibility is restored by introducing the U(1) gauge interaction.  Regarding the dark energy scenario, either the authors of these works did not consider it or the introduction of the U(1) interaction precluded such a possibility.  We acknowledge M. Rinaldi for bringing these papers to our attention.} 

In \cite{Rinaldi:2015iza}, the same author of \cite{Rinaldi:2014yta} studied the Einstein Yang-Mills Higgs (EYMH) model, which is essentially the Higgs dark energy model but this time in the SO(3) representation instead of the SU(2) one, concluding that the Higgs field can also support the late accelerated expansion of the Universe. Although the author did take into account the interaction between the Higgs field and the gauge vector fields, we are going to show in this work that it is precisely this interaction term that plagues the EYMH model with several inconsistencies with the symmetries of the FLRW spacetime which cannot be avoided by fixing the gauge of the Higgs field as in \cite{Alvarez:2019ues}. Nevertheless, although the EYMH model is not compatible with the FLRW universe, it could fit in a homogeneous but anisotropic Bianchi-I background. We expected a non-negligible contribution of the spatial shear to the dark energy content due to the vector nature of the theory, as has been shown in other works where vector fields, gauge fields or p-forms have been used \cite{Koivisto:2008ig, Adshead:2018emn, Almeida:2019iqp}. In this paper, we are going to show that the final state of the Universe is an eternal isotropic accelerated expansion; however, the anisotropic shear today can get appreciable values which can fit in the data analysis \cite{Campanelli:2010zx, Akarsu:2019pwn,Amirhashchi:2018nxl,Saadeh:2016sak} or that could be observed by future missions like ESA's Euclid \cite{Amendola:2016saw}.\footnote{The same conclusion has been reached in Ref. \cite{Paliathanasis:2020pax} in the framework of the Einstein-aether scalar field theory.}

This paper is organized as follows. In section \ref{Incompatibility of the EYMH Model with an Isotropic Universe} we are going to present the general EYMH model in the SO(3) representation and show explicitly the inconsistencies of the model with the FLRW background. Section \ref{Anisotropic Model} is dedicated to the anisotropic EYMH model where we are going to show that, although the only attractor of the theory corresponds to an isotropic accelerated expansion, the current value of the spatial shear can take non-negligible values; the behaviour of the equation of state for dark energy and its possible consequences in the large-scale structure formation process are also going to be shown in this section. In section \ref{An Isotropic Possibility: The Higgs Triad}, we are going to present a way to override the inconsistencies between the EYMH model and the FLRW universe by employing a set of three Higgs triplets that we call the ``\emph{Higgs triad}". Our conclusions are going to be given in section \ref{Conclusions}.

\section{Incompatibility of the EYMH Model with an Isotropic Universe} \label{Incompatibility of the EYMH Model with an Isotropic Universe}

\subsection{The General Model}

The EYMH model involves a Higgs field interacting with a gauge field whose dynamics is given by a Yang-Mills term.  Both, the Higgs field and the gauge fields are  minimally coupled to the gravity which, in turn, is described by the Einstein's General Relativity theory. The EYMH action, in the SO(3) representation, is\footnote{Greek indices run from 0 to 3 and denote space-time components, and Latin indices run from 1 to 3 and denote either SO(3) gauge components or spatial components.} 
\begin{align} \label{EYMH action}
S =& \int  \text{d}^4 x \sqrt{- \text{det} g_{\mu\nu}} \Big[ \frac{m_\text{P}^2}{2} R - \frac{1}{4} F^a_{\ \mu\nu}F_a^{\ \mu\nu} \nonumber \\
&- \frac{1}{2} D_\mu \mathcal{H}^a D^\mu \mathcal{H}_a - V\left(\mathcal{H}^2\right) + \mathcal{L}_{m}  + \mathcal{L}_{r} \Big] \,,
\end{align}
where $g_{\mu\nu}$ is the spacetime metric, $m_\text{P}$ is the reduced Planck mass, $R$ is the Ricci scalar, $F^a_{\ \mu\nu}$ is the SO(3) gauge field strength tensor given by
\begin{equation}
F^a_{\ \mu\nu} \equiv \nabla_\mu A^a_{\ \nu} - \nabla_\nu A^a_{\ \mu} + g \varepsilon^a_{\ b c} A^b_{\ \mu} A^c_{\ \nu} \,,
\end{equation}
$A_\mu^a$ representing the gauge fields,
$g$ being the SO(3) group coupling constant, $\varepsilon^{a}_{\ bc}$ being the Levi-Civita symbol, $D_\mu$ being the SO(3) gauge covariant derivative defined by
\begin{equation}
D_\mu \mathcal{H}^a \equiv \nabla_\mu \mathcal{H}^a + g \varepsilon^a_{\ b c} A^b_{\ \mu} \mathcal{H}^c \,,
\end{equation}
which is applied to the Higgs triplet $\mathcal{H}$:
\begin{equation}
\mathcal{H} \equiv (\mathcal{H}_1 \ \ \mathcal{H}_2 \ \ \mathcal{H}_3)^{\text{T}}  \,,
\end{equation} 
where $\mathcal{H}^a = \mathcal{H}_a$ are real scalar fields, $V(\mathcal{H}^2)$ is the symmetry breaking potential
\begin{equation}
V(\mathcal{H}^2) \equiv \frac{\lambda}{4} \left( \mathcal{H}^2 - \mathcal{H}^2_0 \right)^2 \,,
\end{equation}
with $\lambda$ being the quartic coupling constant, $\mathcal{H}_0$ being the Higgs vacuum value, and $\mathcal{L}_m$ and $\mathcal{L}_r$ being the Lagrangian densities for perfect fluids of matter and radiation respectively.

In the following subsections, we are going to assume that $g_{\mu\nu}$ corresponds to the homogeneous and isotropic FLRW metric and develop the equations of motion for this model employing two different formalisms.

\subsection{The Reduced Action} \label{The Reduced Action}

The symmetries of the spacetime are encoded in the metric. Once these symmetries have been identified, for instance homogeneity and isotropy in a FLRW universe, the matter content has to obey the same symmetries in order to have consistent equations. This consistency between geometry and matter allows one to identify the non-dynamical degrees of freedom in the Lagrangian describing the theory, resulting in a ``smaller" physical phase space. This Lagrangian written in terms of only the dynamical degrees of freedom yields to what is known as the ``reduced action" \cite{Cendra2001}.

The EYMH action in the SO(3) representation, as a mechanism for the late accelerated expansion, was originally proposed in \cite{Rinaldi:2015iza} (see \cite{Moniz:1991kx,Ochs:1996yr} for primordial inflation). In that work, the author employed the reduced theory and a dynamical system analysis which allowed him to conclude that the only attractor of the theory is characterized by a dark energy dominance. In what follows, we have used the reduced action formalism to get the equations of motion of the model, thus reproducing some of the computations worked out in \cite{Rinaldi:2015iza}.

Since the Universe is highly homogeneous, isotropic and spatially flat \cite{Aghanim:2018eyx}, it is assumed that the background is described by the FLRW metric
\begin{equation}
\text{d} s^2 \equiv - N^2(t) \text{d} t^2 + a^2(t) \delta_{i j} \text{d} x^i \text{d} x^j \,,
\end{equation}
where $a(t)$ is the scale factor, and $N(t)$ is the lapse function, both in terms of the cosmic time $t$. As shown in \cite{Bento:1992wy} (see also \cite{ArmendarizPicon:2004pm}), an ansatz for the gauge vector fields obeying the symmetries of this spacetime is
\begin{equation} \label{ansatz gauge}
A^a_0(t) \equiv 0 \,,\quad A^a_{\ i} \equiv a(t) \psi(t) \delta^a_{\ i} \,,
\end{equation}
where $\psi(t)$ is a scalar field. As a shorthand notation we define $\phi(t) \equiv a(t) \psi(t)$. By replacing this ansatz, the Higgs triplet and the FLRW metric in the action (\ref{EYMH action}), we get:
\begin{align} \label{reduced action}
S_{\text{red}} &= \int \text{d} t N a^3 \Big[ \frac{m_\text{P}^2}{2} R(N, a) + \frac{3}{2} \frac{\dot{\phi}^2}{N^2 a^2} - \frac{3}{2} \frac{g^2 \phi^4}{a^4}  \nonumber \\
&+ \frac{1}{2} \frac{\dot{\mathcal{H}}^2}{N^2} - \frac{g^2 \phi^2 \mathcal{H}^2}{a^2} - V(\mathcal{H}^2) + \mathcal{L}_{r} + \mathcal{L}_{m} \Big] \, ,
\end{align}
which corresponds to the reduced action of the model as presented in \cite{Rinaldi:2015iza}.

Varying (\ref{reduced action}) with respect to $N(t)$ we have obtained the first Friedmann equation:\footnote{Where we assume $N(t) = 1$, once the variational principle is applied.}
\begin{equation} \label{reduced H2}
3 m_\text{P}^2 H^2 = \frac{3}{2} \frac{\dot{\phi}^2}{a^2} + \frac{3}{2} \frac{g^2 \phi^4}{a^4} + \frac{1}{2}\dot{\mathcal{H}}^2  + \frac{g^2 \phi^2 \mathcal{H}^2}{a^2} + V + \rho_m + \rho_r ,
\end{equation}
where $\rho_m$ and $\rho_r$ are the densities for matter and radiation, respectively. The second Friedmann equation comes after variation of (\ref{reduced action}) with respect to $a(t)$:
\begin{equation} \label{reduced Hdot}
- 2 m_\text{P}^2 \dot{H} = 2 \frac{\dot{\phi}^2}{a^2} + 2\frac{g^2 \phi^4}{a^4} + \dot{\mathcal{H}}^2 + 2 \frac{g^2 \phi^2 \mathcal{H}^2}{3 a^2} + \rho_m + \frac{4}{3}\rho_r \,.
\end{equation}

The equations of motion for the gauge field and the Higgs field are obtained upon variation of (\ref{reduced action}) with respect to $\phi(t)$ and $\mathcal{H}_a (t)$:
\begin{eqnarray} 
\ddot{\phi} + H \dot{\phi} + \frac{2 g^2 \phi^3}{a^2} + \frac{2 g^2 \mathcal{H}^2 \phi}{3} &=& 0 \,, \label{reduced gauge} \\
\ddot{\mathcal{H}}^a + 3H \dot{\mathcal{H}}^a + \frac{2 g^2 \phi^2 \mathcal{H}^a}{a^2} + \frac{\text{d} V}{\text{d} \mathcal{H}_a} &=& 0 \,,  \label{reduced Higgs}
\end{eqnarray}
respectively. In principle, Eqs. (\ref{reduced H2}) - (\ref{reduced Higgs}) describe the dynamics of the system in a proper way;  we are going to show in the next subsection, however, that this model is indeed incompatible with a FLRW spacetime because the Higgs triplet cannot obey the same symmetries of this spacetime.

\subsection{Einstein Equations and Inconsistencies of the Model} \label{EYMH problems}

In general, the dynamics of the background universe (geometry $+$ matter content) is encoded in the Einstein equations. In principle, this set of equations exhibits ten degrees of freedom; however, some of them are non dynamical and yield redundant equations of motion. It is possible to find the true dynamical degrees of freedom by eliminating the non-dynamical ones in all the equations, thus removing the redundancies \cite{Weinberg:1972kfs}. As it is well known, the Einstein equations read
\begin{equation} \label{Einstein Eqs}
m_\text{P}^2 G_{\mu\nu} = T_{\mu\nu} \,,
\end{equation}
where $G_{\mu\nu}$ is the Einstein tensor and $T_{\mu\nu}$ is the momentum-energy tensor, which is defined as
\begin{equation} \label{energy tensor}
T_{\mu\nu} \equiv g_{\mu\nu} \mathcal{L}_{\text{mat}}  - 2 \frac{\delta \mathcal{L}_{\text{mat}}}{\delta g^{\mu\nu}} \,,
\end{equation}
where $\mathcal{L}_{\text{mat}}$ is the Lagrangian density describing the energy budget filling the Universe. In our case
\begin{align}
\mathcal{L}_{\text{mat}} \equiv &- \frac{1}{4} F^a_{\ \mu\nu}F_a^{\ \mu\nu} - \frac{1}{2} D_\mu \mathcal{H}^a D^\mu \mathcal{H}_a \nonumber \\
&- V\left(\mathcal{H}^2\right) + \mathcal{L}_{m}  + \mathcal{L}_{r} \,.
\end{align}
The model is consistent whenever $G_{\mu\nu}$ and $T_{\mu\nu}$ enjoy the same group of symmetries. Clearly, the first term of the right member of (\ref{energy tensor}) exhibits the same symmetries of the metric $g_{\mu\nu}$. However, it is not obvious whether the second term of such a member also exhibits them; we are going to show that this is not the case. 

The second term of the right hand side in (\ref{energy tensor}) is given by
\begin{equation} \label{first term}
2 \frac{\delta \mathcal{L}_{\text{mat}}}{\delta g^{\mu\nu}} = - F^{a}_{\ \mu\rho} F_{a \ \nu}^{\ \rho} - D_\mu \mathcal{H}^a D_\nu \mathcal{H}_a + 2 \frac{\delta(\mathcal{L}_r + \mathcal{L}_r)}{\delta g^{\mu\nu}} \,.
\end{equation}
Replacing $g_{\mu\nu}$ by the FLRW metric, the cosmic triad ansatz (\ref{ansatz gauge}), and the Higgs triplet in  (\ref{first term}), we can see that there are off-diagonal terms in $T_{\mu\nu}$ since
\begin{equation} \label{inconsistency 1}
2 \frac{\delta \mathcal{L}_{\text{mat}}}{\delta g^{\mu\nu}} \Big|_{\mu\nu = 0i } = g \, \phi \, \varepsilon_{a i b} \dot{\mathcal{H}}^a \mathcal{H}^b \,.
\end{equation}
This contribution to the momentum density is the first inconsistency of the model. Also, from (\ref{first term}), we have recognized a second problem
\begin{equation}
2 \frac{\delta \mathcal{L}_{\text{mat}}}{\delta g^{\mu\nu}} \Big|_{\mu\nu = i \neq j } = g^2 \phi^2 \mathcal{H}_i \mathcal{H}_j \,,
\end{equation}
which implies a contribution to the anisotropic stress. These off-diagonal terms in $T_{\mu\nu}$, whose origins are in the interaction between the gauge vector fields and the Higgs triplet, are clearly at odds with the energy tensor of a perfect fluid in the FLRW universe. Moreover, additional difficulties emerge from the equations of motion of the gauge fields. Varying the action (\ref{EYMH action}) with respect to $A^{a}_{\ \mu}$ we have got
\begin{equation} \label{Gauge EOM}
\nabla_\mu F_a^{\ \mu\nu} - g \varepsilon^{b}_{\ c a} F_b^{\ \mu\nu} A^c_{\ \mu} - g \varepsilon^{b}_{\ a c} g^{\mu\nu} D_\mu \mathcal{H}_b \mathcal{H}^c = 0 \,,
\end{equation} 
such that, after having replaced the FLRW metric, the ansatz (\ref{ansatz gauge}) and the Higgs triplet, we have obtained
\begin{align} \label{inconsistency 2}
\ddot{\phi} + H \dot{\phi} + \frac{2 g^2 \phi^3}{a^2} + g^2 \phi \left( \mathcal{H}^2_2 + \mathcal{H}^2_3 \right) &= 0 \,,\, a, i  = 1 \,, \nonumber \\
\ddot{\phi} + H \dot{\phi} + \frac{2 g^2 \phi^3}{a^2} + g^2 \phi \left( \mathcal{H}^2_1 + \mathcal{H}^2_3 \right) &= 0 \,,\, a, i = 2 \,, \nonumber \\
\ddot{\phi} + H \dot{\phi} + \frac{2 g^2 \phi^3}{a^2} + g^2 \phi \left( \mathcal{H}^2_1 + \mathcal{H}^2_2 \right) &= 0 \,,\, a, i = 3 \,,
\end{align}
meaning that the gauge fields $A^a_{\ i} \equiv \phi(t) \delta^a_{\ i}$ do not evolve in the same way, implying that the isotropy in the model is dynamically broken. Again, this problem comes from the interaction between the gauge fields and the Higgs fields.

Similar irregularities were found in the EYMH model charged under the SU(2) group \cite{Alvarez:2019ues,Emoto:2002fb,Hosotani:2002nq}. There, the authors showed that such a model can be compatible with a FLRW spacetime, once the cosmic triad (\ref{ansatz gauge}) is employed, and the gauge of the Higgs doublet is suitable fixed. However, that strategy does not apply here since any gauge for the Higgs triplet yields to trouble with the FLRW universe. For instance, if we choose the gauge where all the components of the Higgs triplet are equal, there is no contribution to the momentum density and each gauge field evolves in the same way, but the contribution to the anisotropic stress does not vanish. 

Although the EYMH model charged under the SO(3) group is incompatible with the FLRW spacetime, this does not rule it out as a possible model for dark energy. A naive solution to avoid the contributions to the momentum density and anisotropic stress is fixing the gauge of the Higgs triplet as
\begin{equation} \label{Higgs gauge}
\mathcal{H}(t) \equiv \left( \Phi(t)\,, \ 0\,, \ 0 \right) \,,
\end{equation}
with $\Phi(t)$ being a scalar field. However, the model is still inconsistent since the diagonal terms in the energy tensor are not equal:
\begin{align} \label{Anisotropy inconsistency}
T_{1 1} &= a^2 \left( \frac{1}{2} \frac{\dot{\phi}}{a^2} + \frac{1}{2}\frac{g^2 \phi^4}{a^4} + \frac{1}{2} \dot{\Phi}^2 - \frac{g^2 \phi^2 \Phi^2}{a^2} - V \right) \,, \nonumber \\
T_{2 2} &= T_{3 3} = a^2 \left( \frac{1}{2} \frac{\dot{\phi}}{a^2} + \frac{1}{2}\frac{g^2 \phi^4}{a^4} + \frac{1}{2} \dot{\Phi}^2 - V \right) \,;
\end{align}
of course, the interaction term is the responsible for the different components. 

As shown above, the gauge where the Higgs triplet has only one component generates anisotropy but no other undesirable effect. This anisotropy perfectly fits in an axially symmetric Bianchi-I universe as seen in (\ref{Anisotropy inconsistency}). Therefore, we turn our attention to consider the EYMH model, in the SO(3) representation, in an anisotropic Universe and search for the conditions under which non-negligible spatial shear contributions to the late accelerated expansion are obtained. 

\section{Anisotropic Model} \label{Anisotropic Model}

\subsection{Equations of Motion}

Consider the axially symmetric Bianchi-I metric
\begin{equation} \label{Bianchi I}
\text{d} s^2 = - \text{d} t^2 + e^{2\alpha (t)} \left[ e^{-4 \sigma (t)} \text{d} x^2 + e^{2 \sigma (t)} \left( \text{d} y^2 + \text{d} z^2 \right) \right] \,,
\end{equation}
where $a(t) \equiv e^{\alpha (t)}$ is the average scale factor and $\sigma (t)$ is the spatial shear. To be consistent with the symmetries of the metric (\ref{Bianchi I}), the ansatz (\ref{ansatz gauge}) is replaced by a more suitable axially symmetric ansatz \cite{Murata:2011wv}. We have chosen $A^a_{\ 0} (t) \equiv 0$ and
\begin{equation} \label{Axial ansatz}
 A^1_{\ 1} (t) \equiv I (t) \,,\quad A^2_{\ 2} (t) = A^3_{\ 3} (t) \equiv J (t) \,,
\end{equation}
while the other components of the fields are set to zero.\footnote{Other possible ansatz have been employed in \cite{Maleknejad:2011jr, Maleknejad:2013npa, Adshead:2018emn}.} We have also used the shorthand notation
\begin{equation}
G_1 \equiv \sqrt{g^{11}} , \, \, G_2 \equiv \sqrt{g^{22}} = \sqrt{g^{33}} \,. 
\end{equation}
The gauge of the Higgs field is fixed as in (\ref{Higgs gauge}). As mentioned before, this choice assures no contributions either to the momentum density or to the anisotropic stress, while generating two different pressures.

Using the configurations for the Higgs Fields (\ref{Higgs gauge}) and the gauge fields (\ref{Axial ansatz}), together with the Einstein equations (\ref{Einstein Eqs}), we can write the first Friedmann equation ($m_\text{P}^2 G_{00} = T_{00}$) as  
\begin{equation} \label{Ani H2}
3 m_\text{P}^2 H^2 = \rho_r + \rho_m + \rho_{\text{DE}} \,,
\end{equation}
where 
\begin{equation}
\rho_{\text{DE}} \equiv \rho_{\text{YM}} + \rho_{\mathcal{H}} + 3 m_\text{P}^2 \dot{\sigma}^2 \,,
\end{equation}
 is the density of the dark energy split into the contributions from the Yang-Mills term $\rho_{\text{YM}}$ and the Higgs field $\rho_{\mathcal{H}}$.  The latter are given by
\begin{align}
\rho_{\text{YM}} &\equiv \frac{1}{2} (G_1 \dot{I})^2 + (G_2 \dot{J})^2 + g^2 (G_1 I)^2 (G_2 J)^2  \nonumber \\
 &+ \frac{1}{2} g^2 (G_2 J)^4 \,, \\
 \nonumber \\
\rho_{\mathcal{H}} &\equiv \frac{1}{2} \dot{\Phi}^2 + g^2 \Phi^2 (G_2 J)^2 + V \,.
\end{align}

On the other hand, for the second Friedmann equation \mbox{($m_\text{P}^2 \, \text{Tr}\,(G_{i j}) = \text{Tr}\,(T_{i j})$)} we have got
\begin{equation} \label{Ani Hdot}
- 2 m_\text{P}^2 \dot{H} = 3 m_\text{P}^2 H^2 + \frac{1}{3}\rho_r + p_{\text{DE}} \,,
\end{equation}
where $p_\text{DE}$ is the pressure of the dark energy given by
\begin{equation}
p_\text{DE} \equiv \frac{1}{3} \rho_\text{YM} + \frac{1}{2} \dot{\Phi}^2 - \frac{1}{3} g^2 (G_2 J)^2 \Phi^2 - V + 3 m_\text{P}^2 \dot{\sigma}^2 \,,
\end{equation}
and for the spatial shear ($G^2_{\ 2} - G^1_{\ 1} = (T^2_{\ 2} - T^1_{\ 1}) / m_\text{P}^2$)
\begin{eqnarray} \label{Ani shear}
3 m_\text{P}^2 (\ddot{\sigma} + 3 H \dot{\sigma}) &=& (G_1 \dot{I})^2 - (G_2 \dot{J})^2 +  g^2 (G_2 J)^4 \nonumber \\
&& - g^2 (G_1 I)^2 (G_2 J)^2 + g^2 \Phi^2 (G_2 J)^2 \,. \nonumber \\
\end{eqnarray}

Having replaced the field configurations in (\ref{Gauge EOM}), we have got the equations of motion for the gauge fields
\begin{equation} \label{Ani G1}
0 =   (G_1 \ddot{I}) + (H + 4\dot{\sigma})(G_1 \dot{I}) + 2 g^2 (G_2 J)^2 (G_1 I) \,,
\end{equation}
and
\begin{eqnarray}  \label{Ani G2}
0 &=& (G_2 \ddot{J}) + (H - 2\dot{\sigma})(G_2 \dot{J}) + g^2 (G_2 J) (G_1 I)^2 \nonumber \\
&&+ g^2 (G_2 J)^3 + g^2 (G_2 J) \Phi^2  \,.
\end{eqnarray}
Having performed a variation of the action (\ref{EYMH action}) with respect to $\mathcal{H}^a$ and replaced the field configurations, we have got the equation of motion for the Higgs field
\begin{equation} \label{Ani Higgs}
\ddot{\Phi} + 3 H \dot{\Phi} + 2 g^2 (G_2 J)^2 \Phi + \frac{dV}{d\Phi} = 0 \, .
\end{equation}
The set of Eqs. (\ref{Ani H2}), (\ref{Ani Hdot}), (\ref{Ani shear}), (\ref{Ani G1}), (\ref{Ani G2}), and (\ref{Ani Higgs}) describe the cosmological evolution. The asymptotic behaviour of this set of dynamical equations can be studied through a dynamical system analysis \cite{Coley:2003mj, Wainwright2009} (see also Ref. \cite{Alho:2019pku}), this being an approach that we are going to illustrate in the following section.

\subsection{Dynamical System Analysis}

We have introduced the following dimensionless variables
\begin{equation*}
x \equiv \frac{1}{\sqrt{3} m_\text{P}} \frac{(G_1 \dot{I})}{H} \,,\ y \equiv \frac{1}{\sqrt{3} m_\text{P}} \frac{(G_2 \dot{J})}{H} \,,\ z  \equiv \frac{1}{\sqrt{6}} \frac{\dot{\Phi}}{H} \,,
\end{equation*}
\begin{equation*}
w \equiv \frac{1}{\sqrt{3} m_\text{P}} \frac{g (G_2 J) \Phi}{H} \,,\ v \equiv \frac{1}{m_\text{P} H} \sqrt{\frac{\ V}{3}} \,,\ \xi \equiv \frac{\sqrt{3} m_\text{P}}{(G_2 J)} \,,
\end{equation*}
\begin{equation*}
p \equiv \sqrt{\frac{g}{\sqrt{3} m_\text{P} H}} (G_1 I) \,,\ s \equiv \sqrt{\frac{g}{\sqrt{3} m_\text{P} H}} (G_2 J) \,,
\end{equation*}
\begin{equation} \label{variables}
\Sigma \equiv \frac{\dot{\sigma}}{H} \,,\ \Omega_m \equiv \frac{\rho_m}{3 m_\text{P}^2 H^2} \,,\ \Omega_r \equiv \frac{\rho_r}{3 m_\text{P}^2 H^2} \,,
\end{equation}
such that the first Friedmann equation (\ref{Ani H2}) becomes the constraint
\begin{align} \label{Friedmann constraint}
\Omega_r &= 1 - \frac{1}{2}x^2 - y^2 - \frac{1}{2} s^2 \left( s^2 + 2 p^2 \right) - z^2 - v^2 \nonumber \\
 &- w^2 - \Sigma^2 - \Omega_m \,.
\end{align}
The set of background equations of motion (\ref{Ani Hdot}), (\ref{Ani shear}), (\ref{Ani G1}), (\ref{Ani G2}), and (\ref{Ani Higgs}) is replaced by the autonomous set\footnote{A prime represents a derivative with respect to the number of e-folds $N \equiv \text{ln} \, a$.}
\begin{align} \label{Auto set}
x' &= x ( q - 1 - 2\Sigma ) - 2 p \, \xi \, s^3 \,, \\
y' &= y( q - 1 + \Sigma ) - \xi \, ( p^2 s^2 + s^4 + w^2 ) \,, \\
z' &= z( q - 2 ) - w \, \xi \, ( \sqrt{2}s^2 + \alpha v) \,, \\
w' &= w ( q + y \, \xi - \Sigma ) + \sqrt{2} \,\xi \, z s^2 \,, \\
v' &= v ( q + 1 ) + \alpha w z \, \xi \,, \\
\xi' &= \xi \, ( 1 - y \,\xi + \Sigma) \,, \\
p' &= \frac{1}{2} p ( q - 1 + 4\Sigma ) + s x \, \xi \,, \\
s' &= \frac{1}{2} s ( q - 1 + 2 y \, \xi - 2\Sigma) \,, \\
\Sigma' &= \Sigma( q - 2 ) + s^2 ( s^2 - p^2) + w^2 - y^2 + x^2 \,, \\
\Omega_m' &= 2 \Omega_m \left( q - 1/2 \right) \,,
\end{align}
where the deceleration parameter $q$ is given by
\begin{equation}
q \equiv - 1 - \frac{\dot{H}}{H^2} = 1 - 2v^2 - w^2 + z^2 - \frac{1}{2} \Omega_m + \Sigma^2 \,,
\end{equation}
and $\alpha$ is a positive dimensionless constant defined by $\alpha \equiv \sqrt{2 \lambda / g^2 \,}$. The deceleration parameter, or equivalently the effective equation of state $w_{\text{eff}} \equiv (2q - 1) / 3$, characterizes the evolution of the average scale factor $a(t)$.

Although the variables (\ref{variables}) describe the autonomous set, the relevant quantities we have been interested in are the physical fields, i.e. the physical gauge vector fields
\begin{equation}
\psi_1 \equiv \frac{(G_1 I)}{m_\text{P}} = \frac{\sqrt{3} p}{s \, \xi} \,,\quad \psi_2 \equiv \frac{(G_2 J)}{m_\text{P}} = \frac{\sqrt{3}}{\xi} \,,
\end{equation}
their speeds
\begin{align}
\psi_1' &\equiv \frac{(G_1 I)'}{m_\text{P}} = \psi_1 (2 \Sigma - 1) + \sqrt{3} x \,, \nonumber \\
\psi_2' &\equiv \frac{(G_2 J)'}{m_\text{P}} = \psi_2 ( y \, \xi - 1 - \Sigma ) \,,
\end{align}
and the Higgs field and its speed
\begin{equation}
\frac{\Phi}{m_\text{P}} = \frac{\sqrt{3} w}{s^2 \, \xi} \,,\quad \frac{\Phi'}{m_\text{P}} = \sqrt{6} z \,.
\end{equation}
Besides the behaviour of each field, the dark sector as a whole can be characterized by its density parameter $\Omega_{\text{DE}} \equiv \rho_{\text{DE}} / 3 m_{\text{P}} H^2$ and its equation of state $w_{\text{DE}} \equiv p_{\text{DE}} / \rho_{\text{DE}}$.

In the following, we are going to discuss the fixed points relevant to the radiation era ($\Omega_r \simeq 1, w_{\text{eff}} \simeq 1/3$), the matter era ($\Omega_m \simeq 1, w_{\text{eff}} \simeq 0$), and the dark energy era ($\Omega_{\text{DE}} \simeq 1, w_{\text{eff}} < - 1/3$).

\subsubsection{\textbf{Radiation dominance}}
\begin{equation*}
z = 0 \,,\ w = 0 \,,\ v = 0 \,,\ \Sigma = 0 \,,\ \Omega_m = 0 \,,
\end{equation*}
\begin{equation}
\Omega_r = 1 - \Omega_{\text{DE}} \,,\ \Omega_{\text{DE}} = \frac{3}{2} \left( s^4 + x^2 \right) \,.
\end{equation}
For this manifold, $w_\text{eff} = 1/3$, which means radiation domination. It contains several submanifolds where $\Omega_{\text{DE}}$ is either non negligible, corresponding to anisotropic scaling solutions with $w_{\text{DE}} = 1/3$, or zero, corresponding to the usual radiation domination.

\begin{itemize}
\item Anisotropic scaling $(\emph{R-1})$:
\end{itemize}
\begin{equation*}
x^2 = p^2 s^2 + y^2 - s^4 \,,\ \xi = 0 \,,
\end{equation*}
\begin{equation}
\Omega_{\text{DE}} =  \frac{3}{2} \left( p^2 s^2 + y^2 \right) \,.
\end{equation}

\begin{itemize}
\item Anisotropic scaling $(\emph{R-2})$:
\end{itemize}
\begin{equation}
x = y \,,\ s = 0 \,,\ \xi = 0 \,,\ \Omega_\text{DE} = \frac{3 x^2}{2}\,.
\end{equation}

\begin{itemize}
\item Anisotropic scaling $(\emph{R-3})$:
\end{itemize}
\begin{equation}
x = \frac{1}{\xi} \,,\ y = \frac{1}{\xi} \,,\ s = 0 \,,\ \Omega_\text{DE} = \frac{3}{2 \xi^2 }\,.
\end{equation}

These are saddle submanifolds where the dark sector behaves as a radiation fluid since $w_{\text{DE}} = 1 / 3$. From the eigenvalues and eigenvectors, we have concluded that the Higgs field is evolving in a decelerated way since $z = 0$ is a $z$-direction attractor, while $w = 0$ and $v = 0$ are $w$-direction and $v$-direction repellers, respectively, such that the interaction term does not vanish in these submanifolds and the Higgs field has not reached its vacuum value yet. When there is no contribution from the dark sector, $w_{\text{DE}}$ is undetermined and $\Omega_r = 1$, which is the usual isotropic radiation dominance. In this point, $\Sigma = 0$ is a $\Sigma$-direction attractor, while $\Omega_m = 0$ is a $\Omega_m$-direction repeller, meaning that the shear is decaying and the density parameter for matter is growing.

\subsubsection{\textbf{Matter dominance}}

\begin{itemize}
\item Isotropic matter $(\emph{M-1})$:
\end{itemize}
\begin{equation*}
x = 0 \,,\ y = 0 \,,\ \xi = 0 \,,\ p = 0 \,,\ s = 0 \,,\ z = 0 \,, 
\end{equation*}
\begin{equation}
w = 0 \,,\ v = 0 \,,\ \Sigma = 0 \,,\ \Omega_m = 1 \,.
\end{equation}
This is a saddle point with $w_{\text{eff}} = 0$, and $w_{\text{DE}}$ undetermined. In consequence, it corresponds to a proper isotropic matter epoch. From the eigensystem analysis we see that $x = 0, y = 0, p = 0, s = 0, z = 0, \Sigma = 0$ and $\Omega_m = 1$ are $x$-direction, $y$-direction, $p$-direction, $s$-direction, $z$-direction, $\Sigma$-direction, and $\Omega_m$-direction attractors respectively. This means that the speeds of the gauge fields, the speed of the Higgs field, and the shear are decaying during the matter epoch, characterized by $\Omega_m \simeq 1$. On the other hand, $w = 0, v = 0$ and $\xi = 0$ are $w$-direction, $v$-direction and $\xi$-direction repellers, respectively, meaning that, during this epoch, the interaction term does not vanish, the Higgs field has not reached its vacuum value yet, and the gauge field $\psi_2$ is decreasing in magnitude.

\subsubsection{\textbf{Dark energy dominance}}

\begin{itemize}
\item Isotropic dark energy $(\emph{DE-1})$:
\end{itemize}
\begin{equation*}
x = 0 \,,\ y = 0 \,,\ \xi = 0 \,,\ p = 0 \,,\ s = 0 \,,\ z = 0 \,, 
\end{equation*}
\begin{equation}
w = 0 \,,\ v = 1 \,,\ \Sigma = 0 \,,\ \Omega_m = 0 \,.
\end{equation}
In this point we have $w_\text{eff} = w_{\text{DE}} = -1$ and $\Omega_\text{DE} = 1$, corresponding to an accelerated expansion solution where the Universe is filled by the Higgs potential. This is a saddle point with only one positive eigenvalue, while the other eigenvalues are negative. The eigenvector associated to this positive eigenvalue indicates that $\xi = 0$ is a repeller in the $\xi$-direction, meaning that the gauge field $\psi_2$ is decreasing in magnitude. Note that $w_{\text{eff}}$ does not depend on $\xi$, consequently $w_\text{eff} = -1$ is an attractor since all the other variables are attracted to their corresponding values in the point. Hence, we say that $(\emph{DE-1})$ is an ``\emph{effective attractor}". Also notice that the shear $\Sigma$ is diluted in the point, i.e. the model does not support an anisotropic dark energy final state. This is consistent with the fact that the gauge fields, $\psi_1$ and $\psi_2$, and their speeds decay to zero in the point, implying that the interaction between the Higgs field and the gauge fields vanishes, and so does the anisotropic shear which is supported by this interaction. We have also been able to infer that the Higgs field $\Phi$ is attracted in a decelerated way (in magnitude) towards a constant value, since $z = 0$ in the point. 

We want to stress that, although the final state of the Universe is characterized by an isotropic accelerated expansion, this does not rule out an observable anisotropic dark energy today. In the following, we are going to present the results of the numerical integration of the autonomous set, in order to see a particular trajectory of the Universe, and look for initial conditions yielding to a significant anisotropic contribution to the dark energy components nowadays.

\subsection{Numerical Solution}

We have performed a numerical integration of the autonomous system. In particular, we have chosen\footnote{This is the order of magnitude obtained when calculating $\alpha$, based on the Standard Model of Particle Physics \cite{Kane:1987gb}.} $\alpha = 1$ and\footnote{Here, the subscript $i$ means that the corresponding quantity is evaluated at some time in the deep radiation epoch.} 
\begin{equation*}
x_i = y_i = \xi_i = 10^{-8} \,,\quad p_i = s_i = 10^{-12} \,,
\end{equation*}
\begin{equation*}
z_i = w_i = 10^{-12} \,,\ v_i = 2 \times 10^{-14} \,,\ \Sigma_i = 10^{-20} \,,
\end{equation*}
\begin{equation} \label{Initial conditions}
\Omega_{m_i} = 4.99 \times 10^{-5} \,,\quad \Omega_{r_i} = 0.99995 \,,
\end{equation}
as initial conditions at redshift $z_r = 6.566 \times 10^{7}$, well within the deep radiation era. The Universe is supposed to have undergone an inflationary phase prior to this epoch, therefore the initial value for the spatial shear is very close to zero. The initial values of the other fields have been chosen so that the dark components provide a subdominant contribution to the radiation era, while obeying the Friedmann constraint (\ref{Friedmann constraint}). In the following, we are going to describe the evolution of the cosmological epochs relevant to the radiation, matter and dark energy dominance eras.

\subsubsection{\textbf{Radiation Dominance}}

\begin{figure}
\subfloat[\label{higgsradiation}]{%
  \includegraphics[height=2.5cm, width=0.48\linewidth]{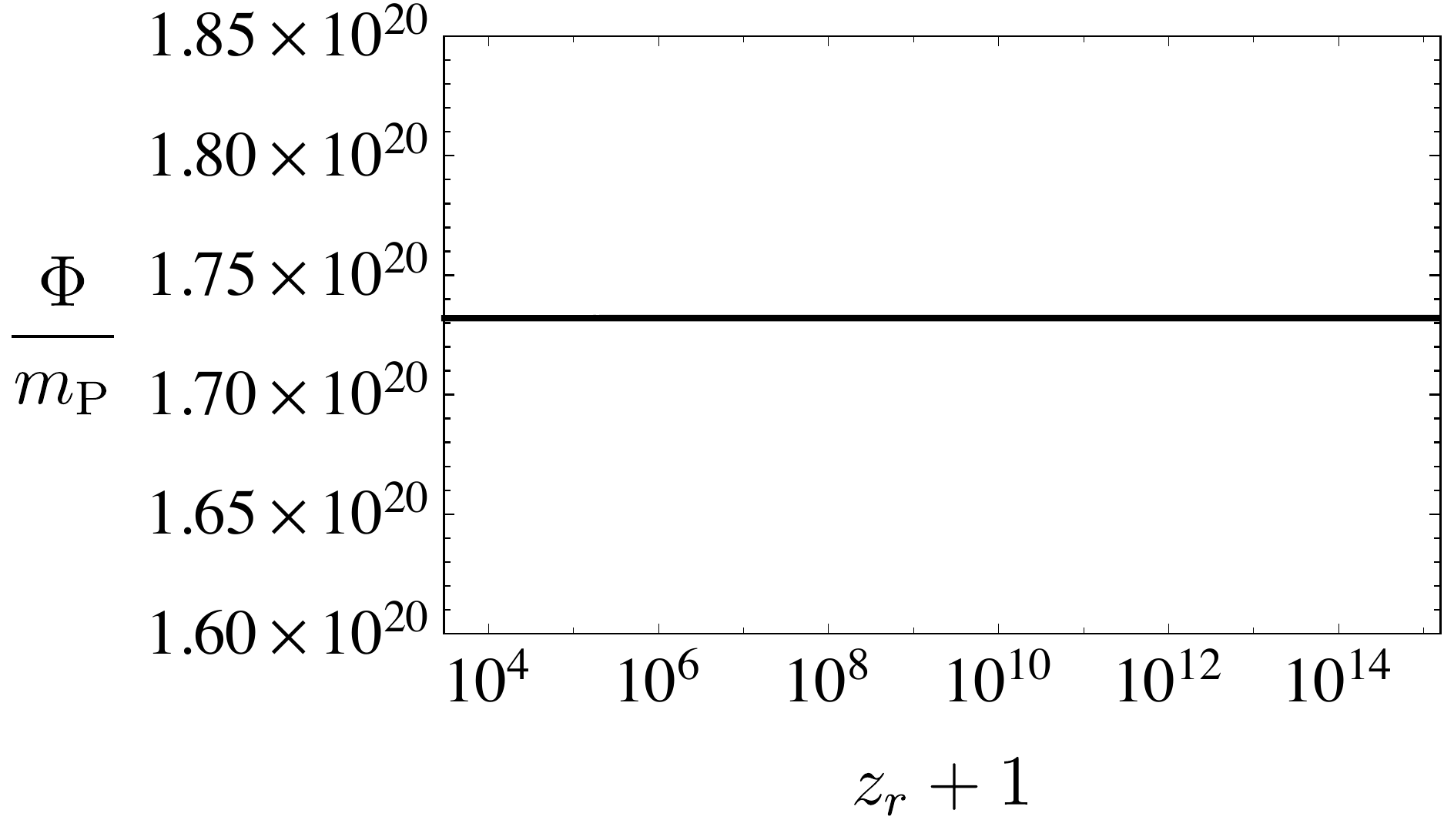}%
}\hfill
\subfloat[\label{higgsspeedradiation}]{%
  \includegraphics[height=2.5cm, width=.48\linewidth]{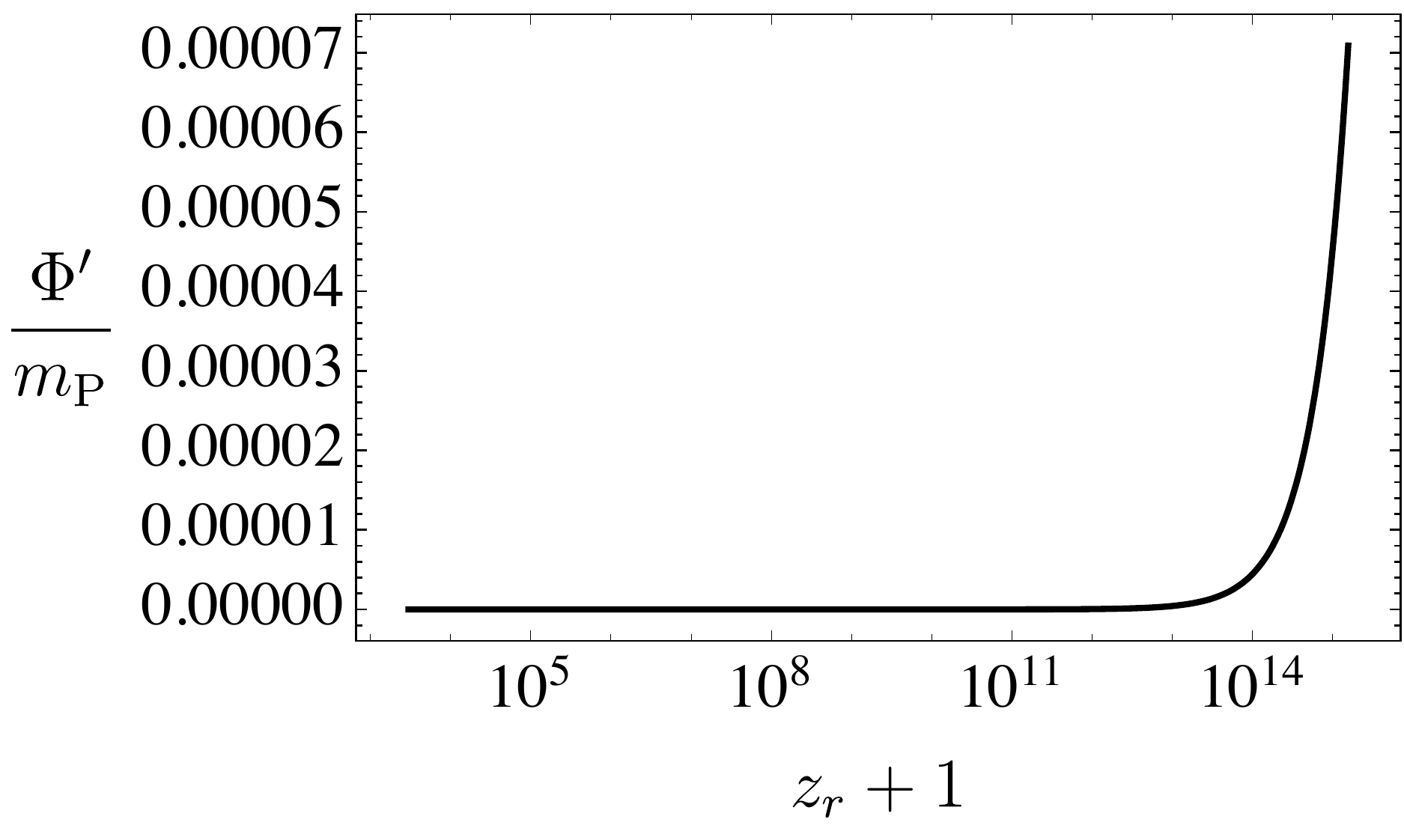}%
}\\
\subfloat[\label{vectorradiation}]{%
  \includegraphics[height=2.5cm, width=.48\linewidth]{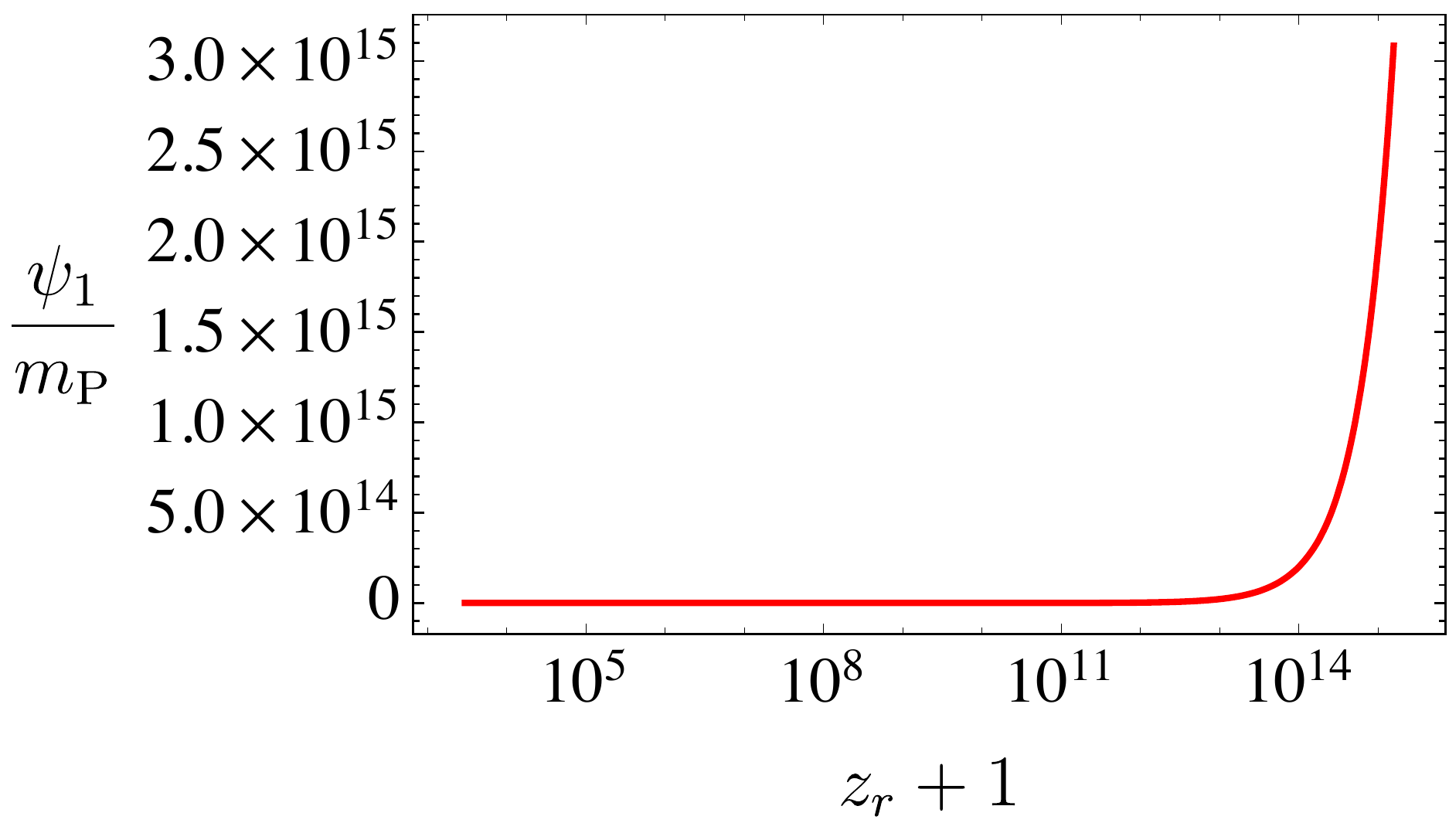}%
}\hfill  
\subfloat[\label{vectorspeedradiation}]{%
  \includegraphics[height=2.5cm,width=.48\linewidth]{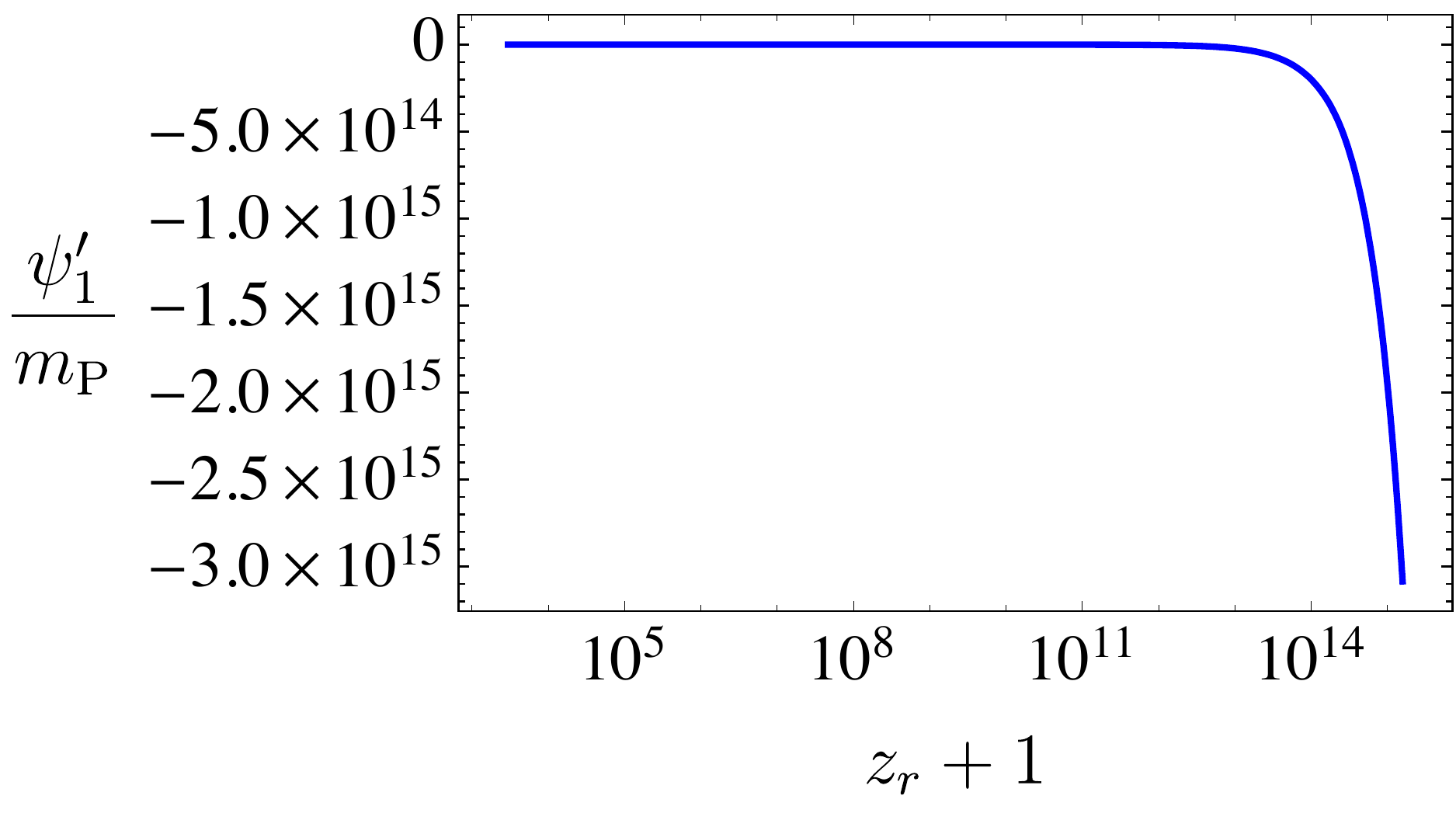}%
}\\
\subfloat[\label{vectorradiation}]{%
  \includegraphics[height=2.5cm,width=.48\linewidth]{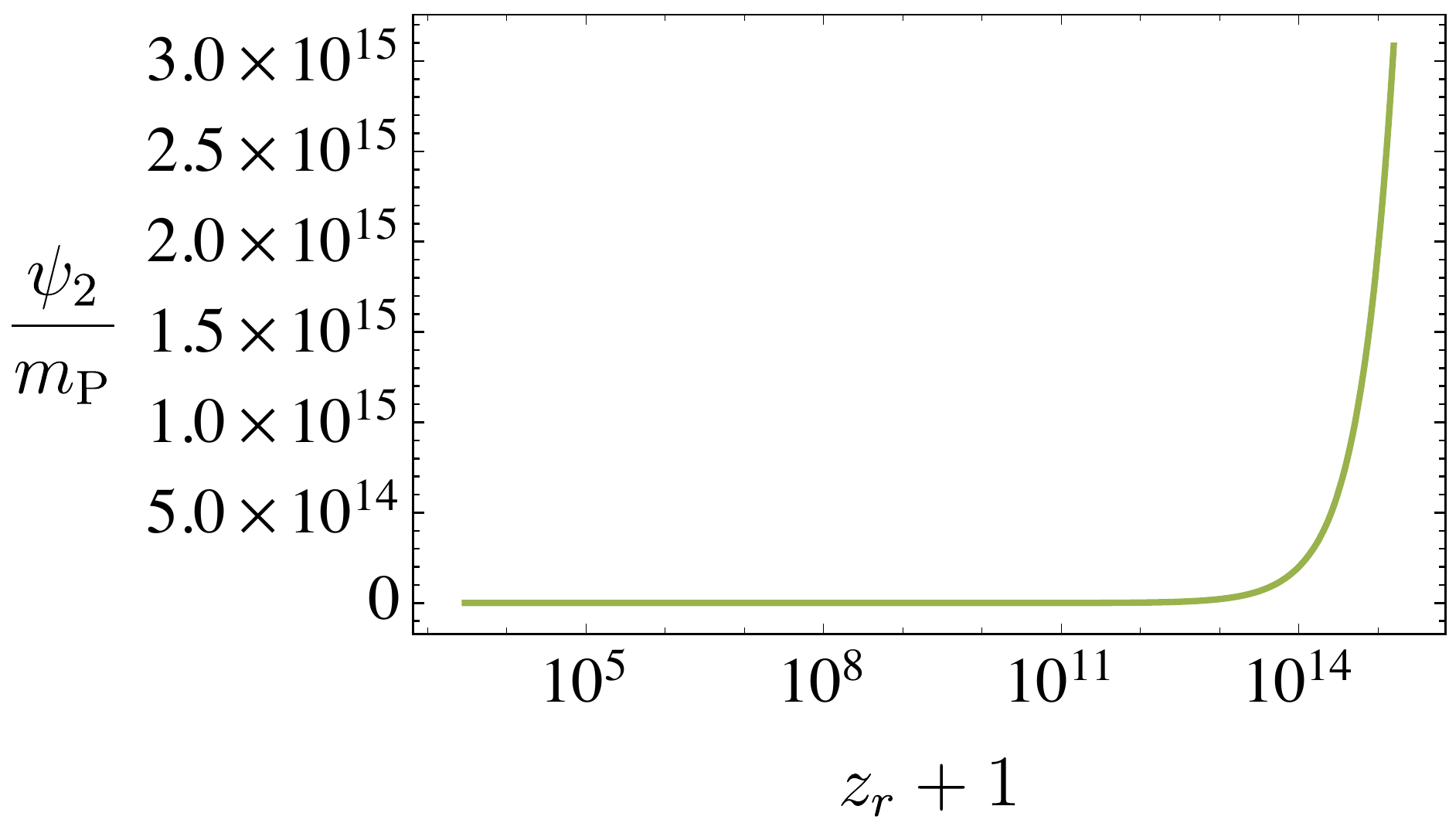}%
}\hfill  
\subfloat[\label{vectorspeedradiation}]{%
  \includegraphics[height=2.5cm,width=.48\linewidth]{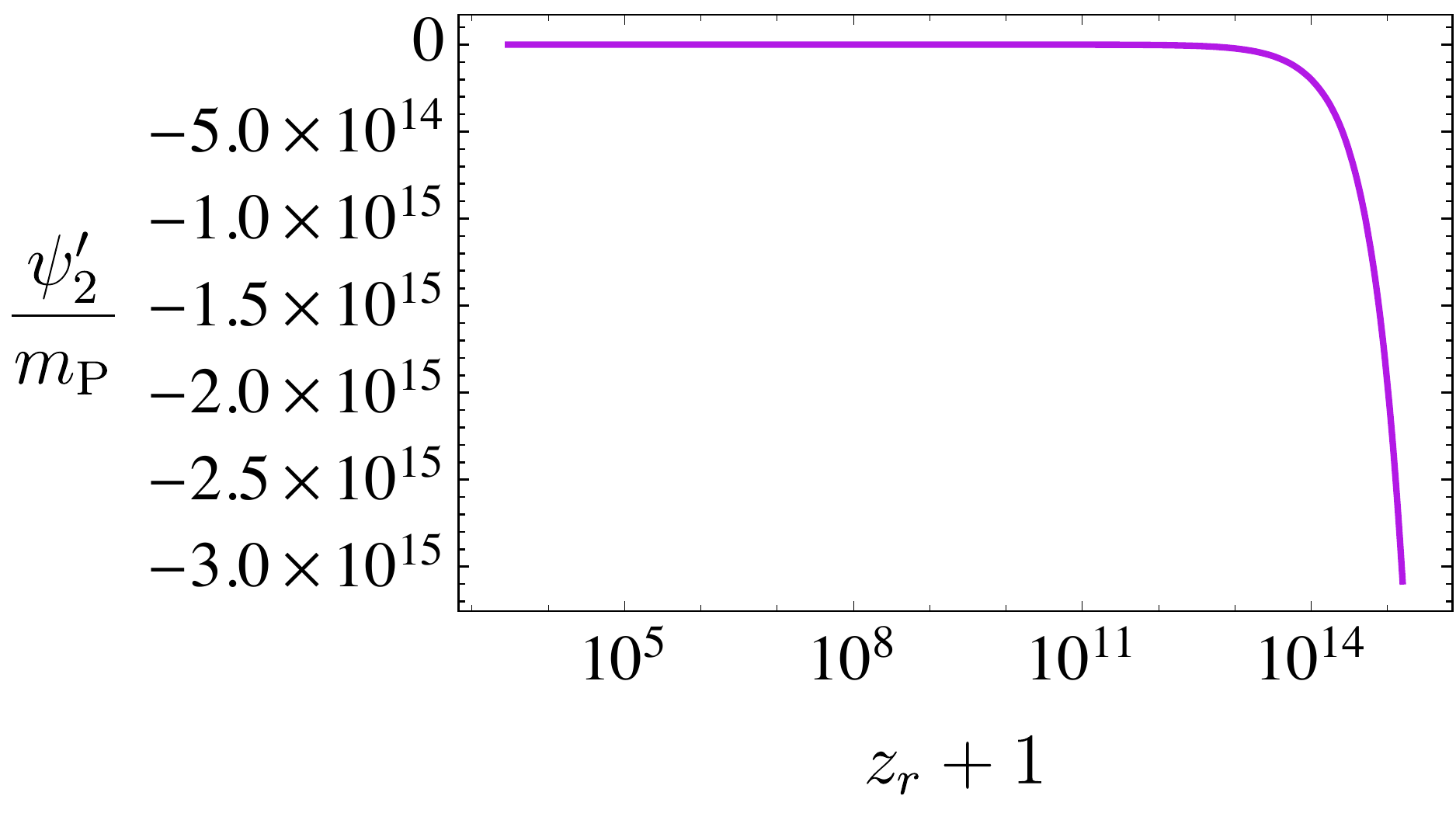}%
}   
\caption{Evolution of the physical fields during the radiation domination epoch. Plots (a) - (f) show the Higgs field, its speed, the physical vector fields and their speeds, respectively. The behaviour is in agreement with that obtained via dynamical systems.}
\label{RadiationFields}
\end{figure}

This period runs from $z_r \approx 10^{15}$ to $z_r \approx 3200$, in agreement with the constraint in the length of the radiation epoch (see Eq. (36) and Appendix B of Ref. \cite{Alvarez:2019ues}).  In Fig. \ref{RadiationFields} we can see that the Higgs field and the gauge fields are evolving in a decelerated way. The magnitude of the Higgs field is almost a constant (its evolution is not appreciated in the plot), and the gauge fields are decaying from a huge value. The contribution of early dark energy is $\Omega_{\text{DE}} \approx 3.5 \times 10^{-15}$ well below the Big-Bang Nucleosynthesis (BBN) constraint $\Omega_\text{DE} < 0.045$ at $z_r = 3200$ \cite{Bean:2001wt}. In this particular case, the main contribution to the dark sector comes from the Higgs field; however, it is possible to find initial conditions where the Yang-Mills term dominates the dark sector during this period.

\subsubsection{\textbf{Matter Dominance}}

\begin{figure}
\subfloat[\label{higgsspeedradiation}]{%
  \includegraphics[height=2.5cm, width=.48\linewidth]{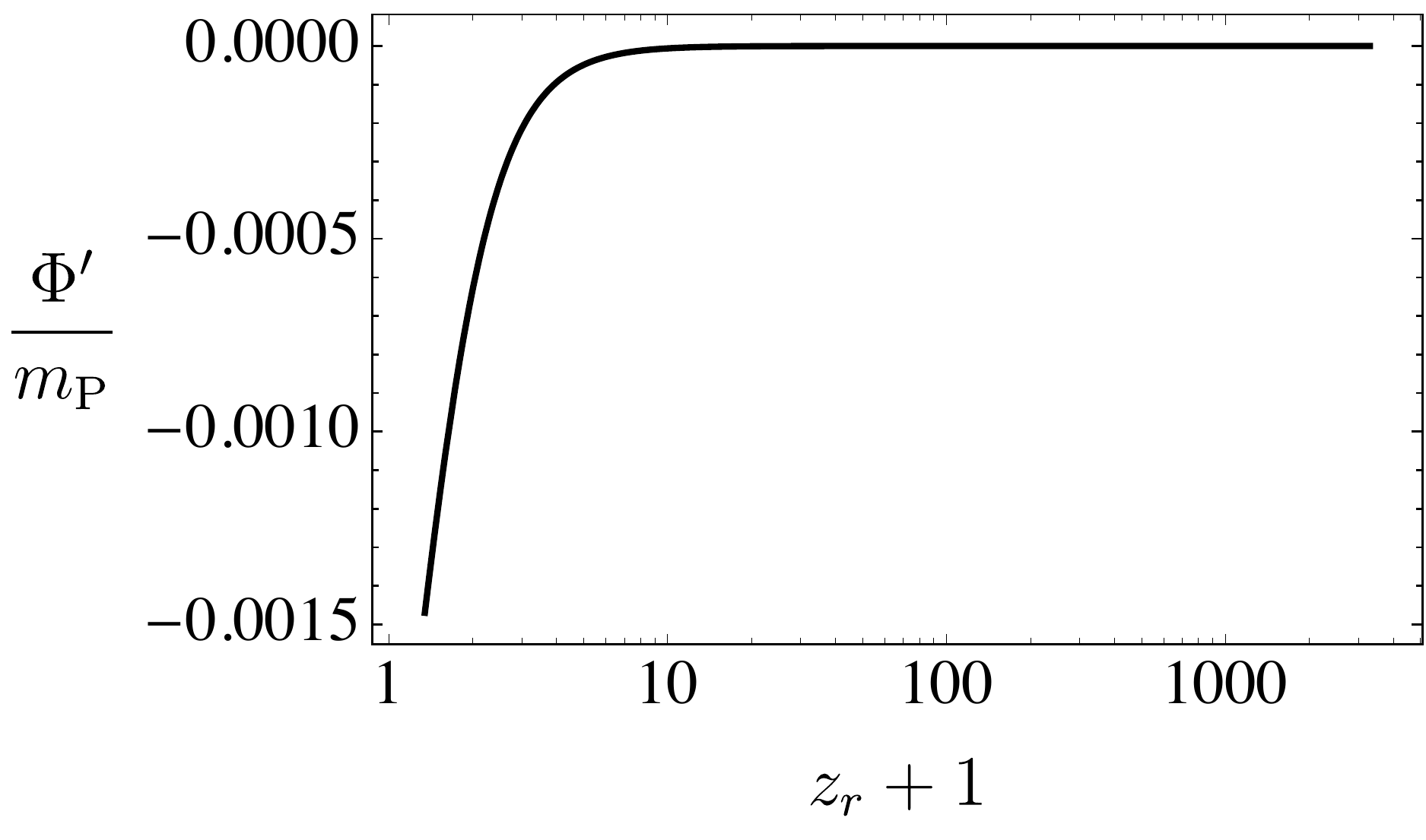}%
}\\
\subfloat[\label{vectorradiation}]{%
  \includegraphics[height=2.5cm, width=.48\linewidth]{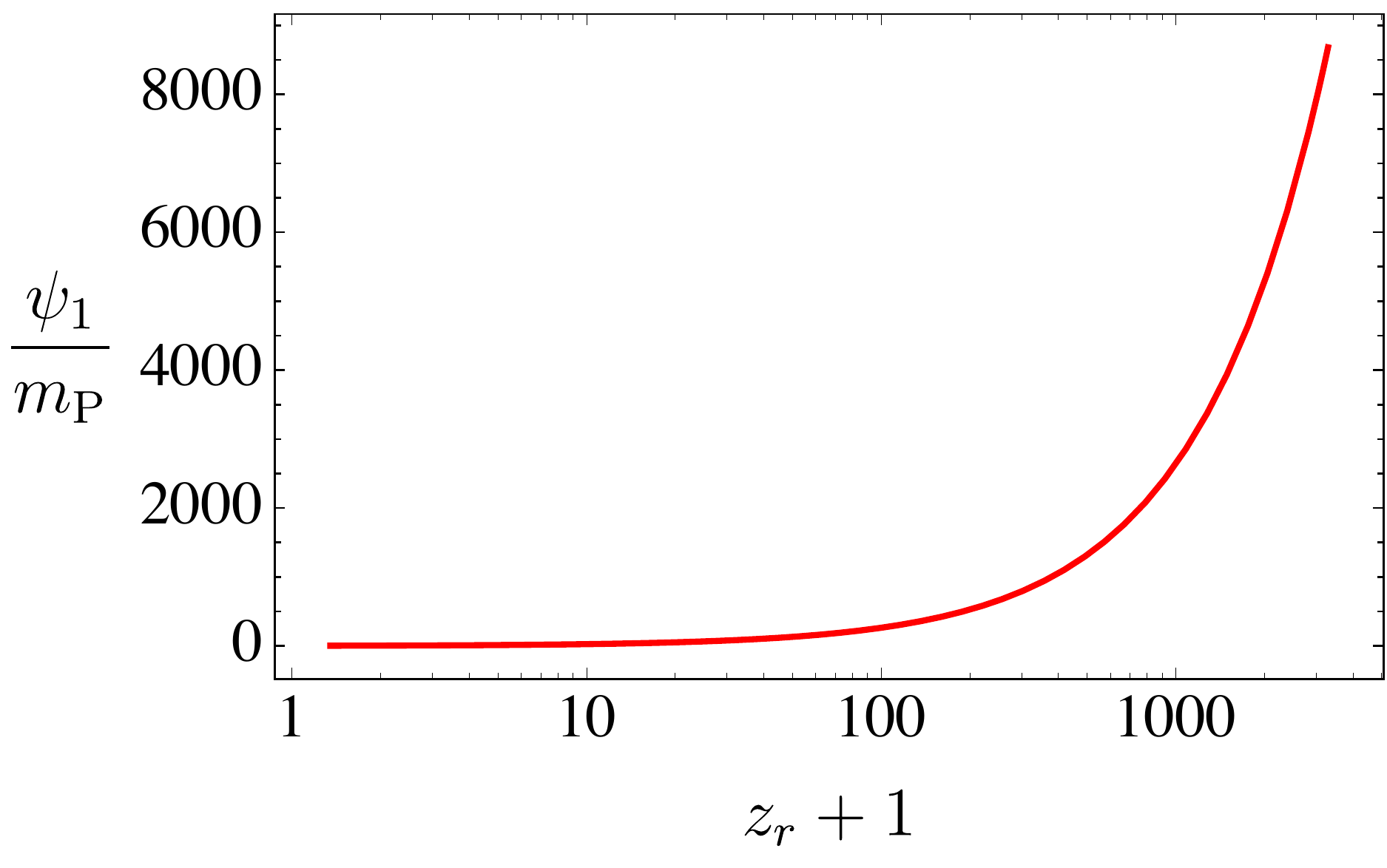}%
}\hfill
\subfloat[\label{vectorspeedradiation}]{%
  \includegraphics[height=2.5cm,width=.48\linewidth]{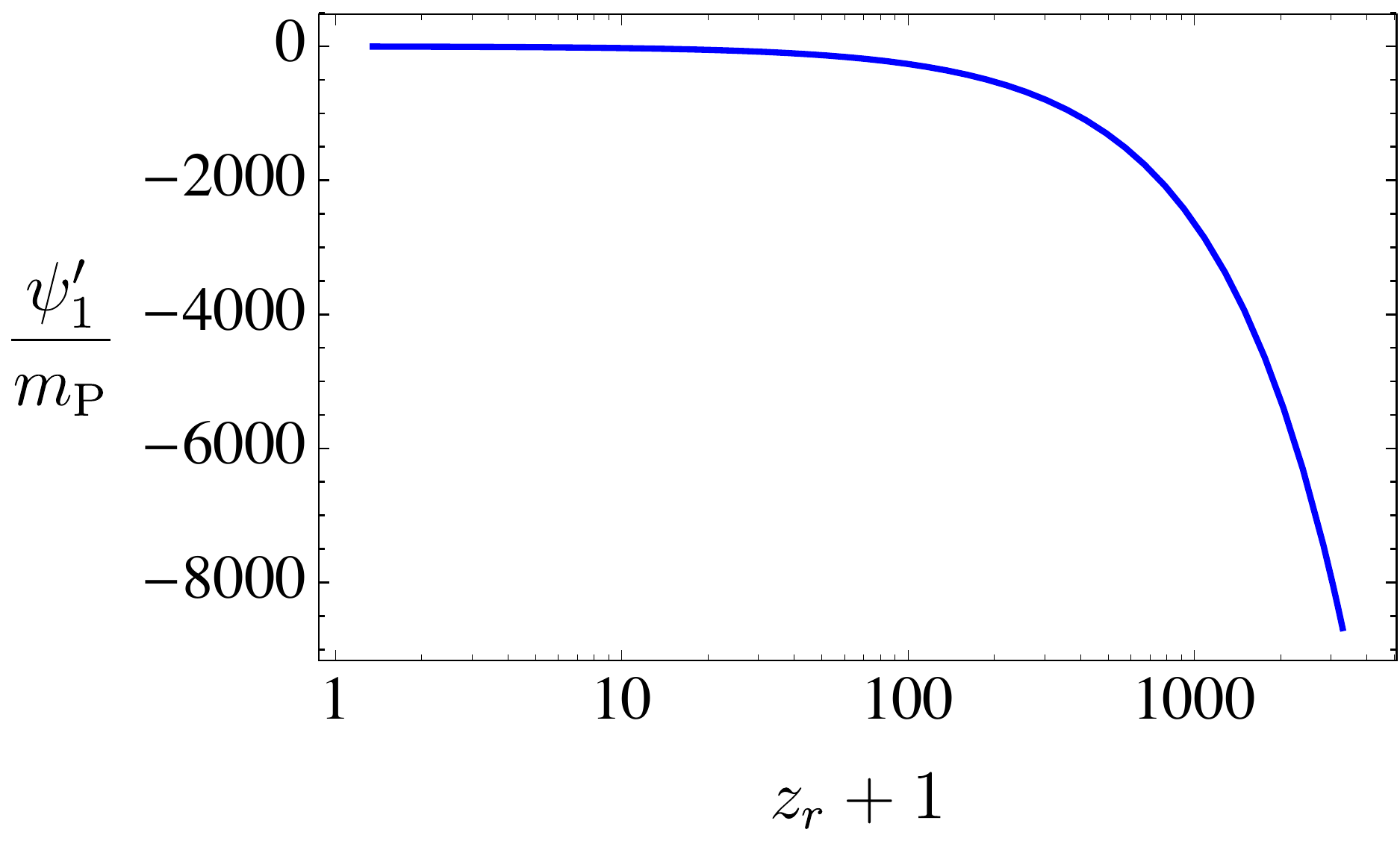}%
}\\
\subfloat[\label{vectorradiation}]{%
  \includegraphics[height=2.5cm,width=.48\linewidth]{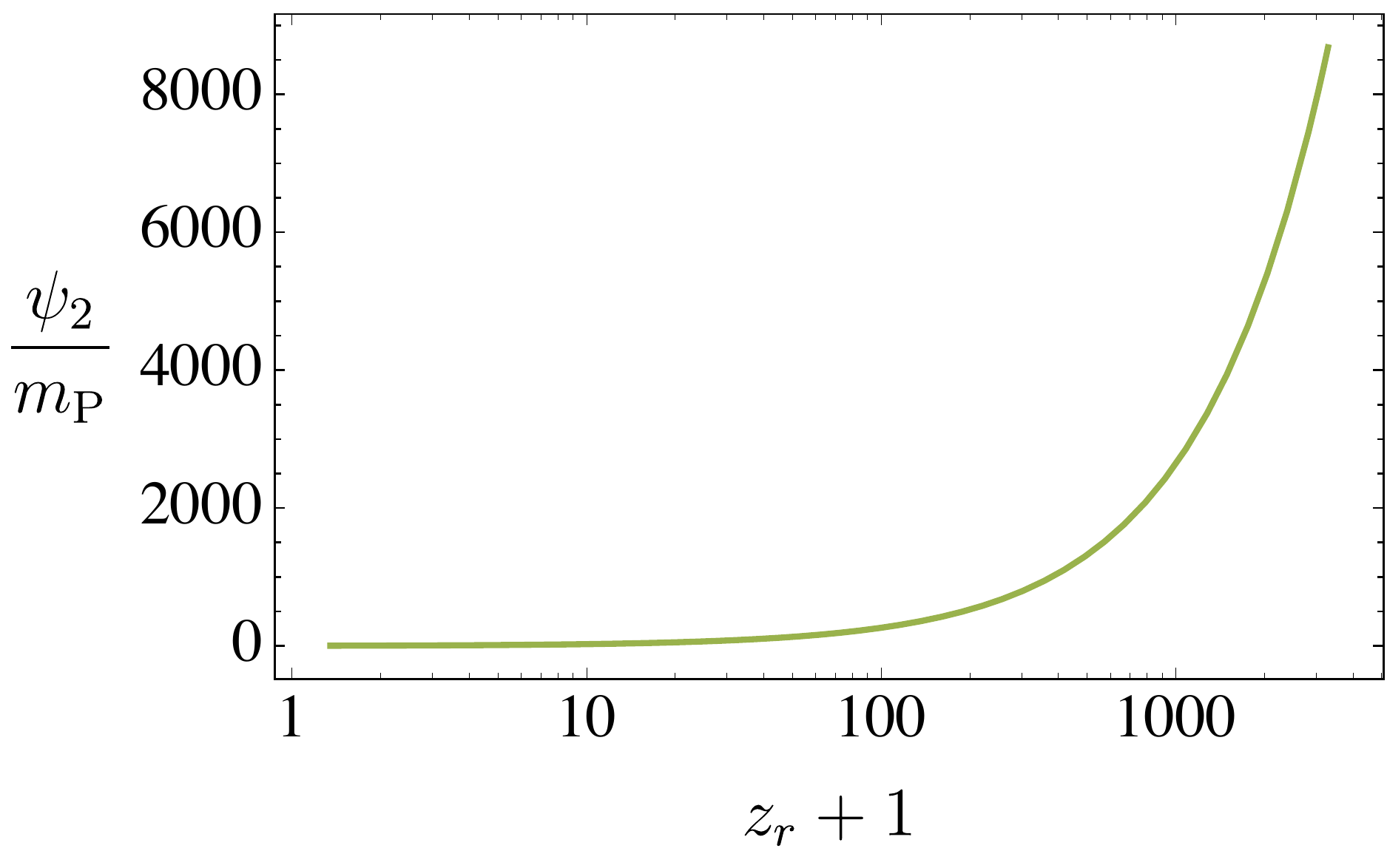}%
}\hfill
\subfloat[\label{vectorspeedradiation}]{%
  \includegraphics[height=2.5cm,width=.48\linewidth]{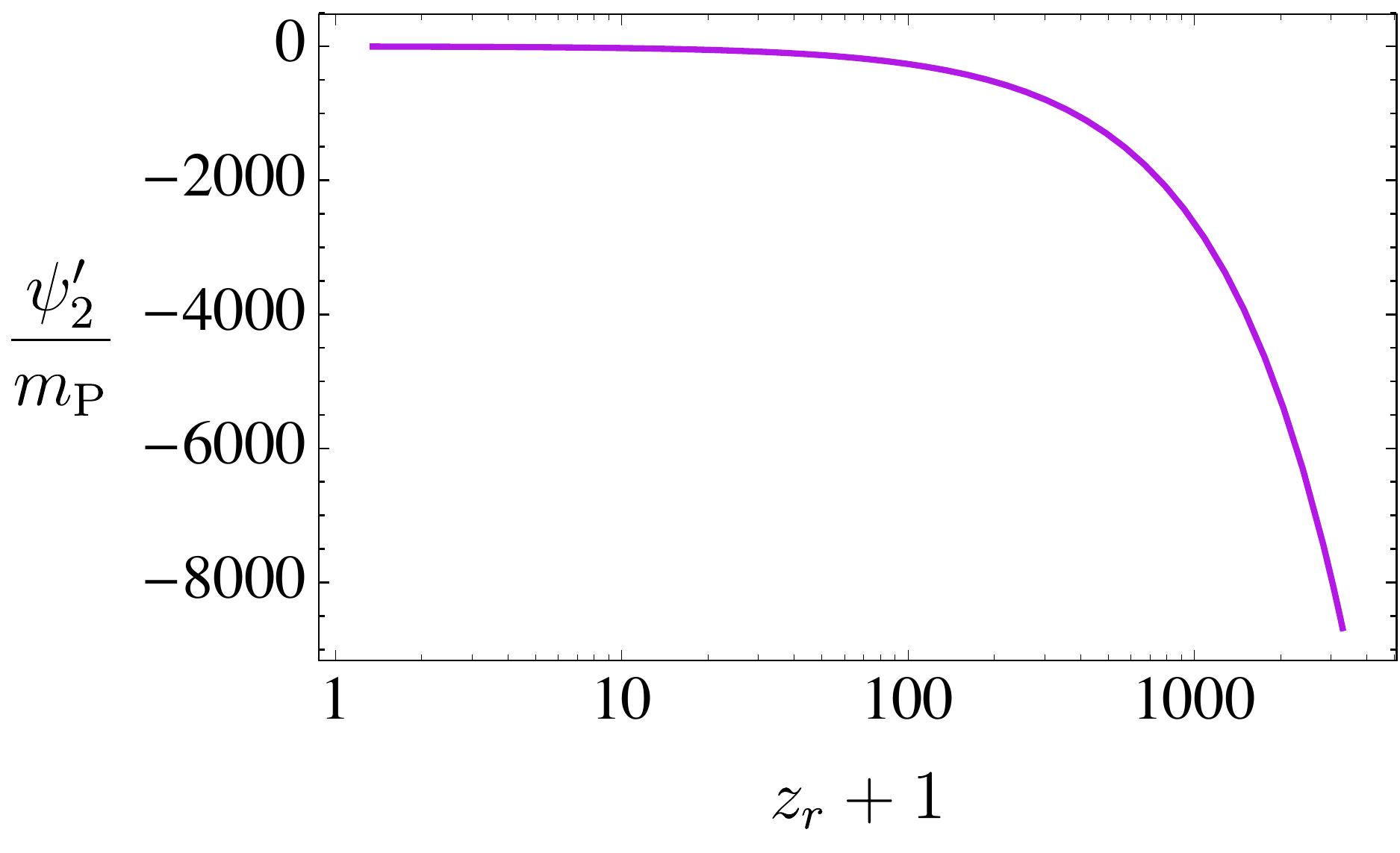}%
}   
\caption{Evolution of the physical fields during the matter domination epoch. Plots a) - e) show the speed of the Higgs field, the physical vector fields and their speeds, respectively. The evolution of the Higgs field is not presented since, by this stage, it has almost reached its asymptotic value. The behaviour is in agreement with that obtained via dynamical systems.}
\label{MatterFields}
\end{figure}

This period runs from $z_r \approx 3200$ to $z_r \approx 0.3$.  In Fig. \ref{MatterFields} we can see that the Higgs field and the gauge fields are still evolving in a decelerated way. The magnitude of the Higgs field is almost a constant and the gauge fields reach values near zero. The contribution of dark energy is $\Omega_{\text{DE}} \approx 1.7 \times 10^{-5}$ well below the CMB constraint $\Omega_\text{DE} < 0.02$ at $z_r = 50$ \cite{Ade:2015rim}.

\subsubsection{\textbf{Dark Energy Dominance}}

\begin{figure}[t]
\subfloat[\label{higgsspeedradiation}]{%
  \includegraphics[height=2.5cm, width=.48\linewidth]{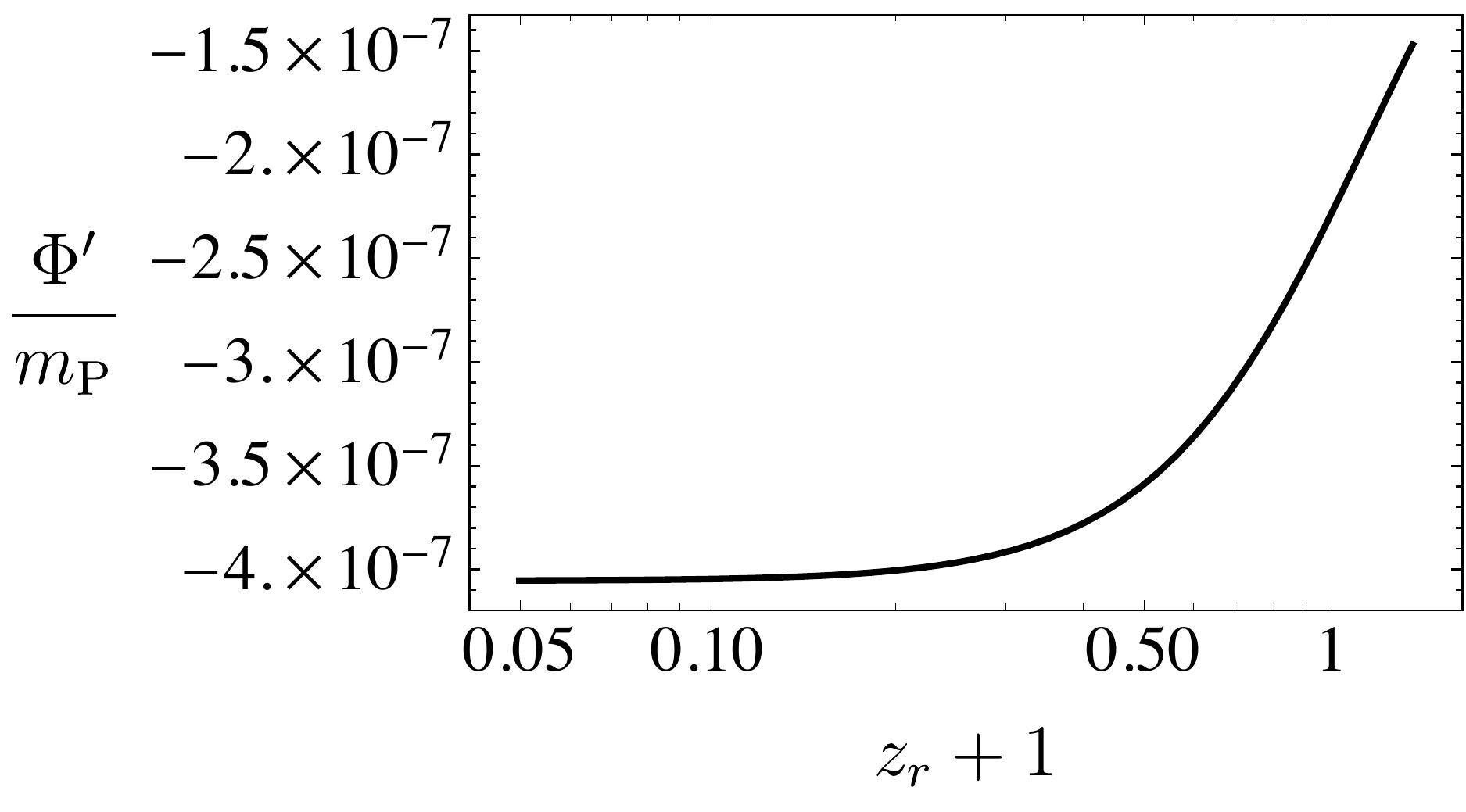}%
}\\
\subfloat[\label{vectorradiation}]{%
  \includegraphics[height=2.5cm, width=.48\linewidth]{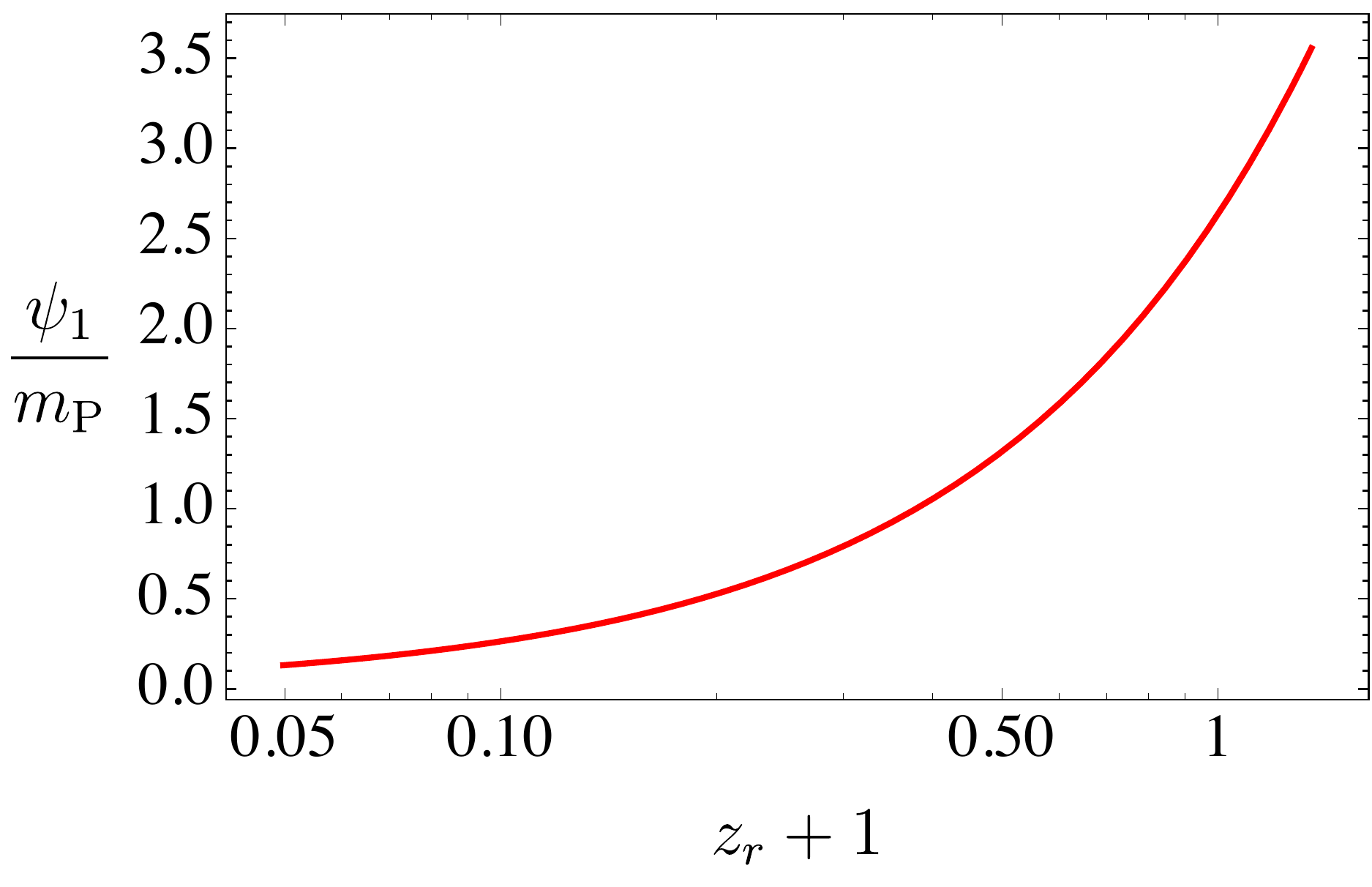}%
}\hfill
\subfloat[\label{vectorspeedradiation}]{%
  \includegraphics[height=2.5cm,width=.48\linewidth]{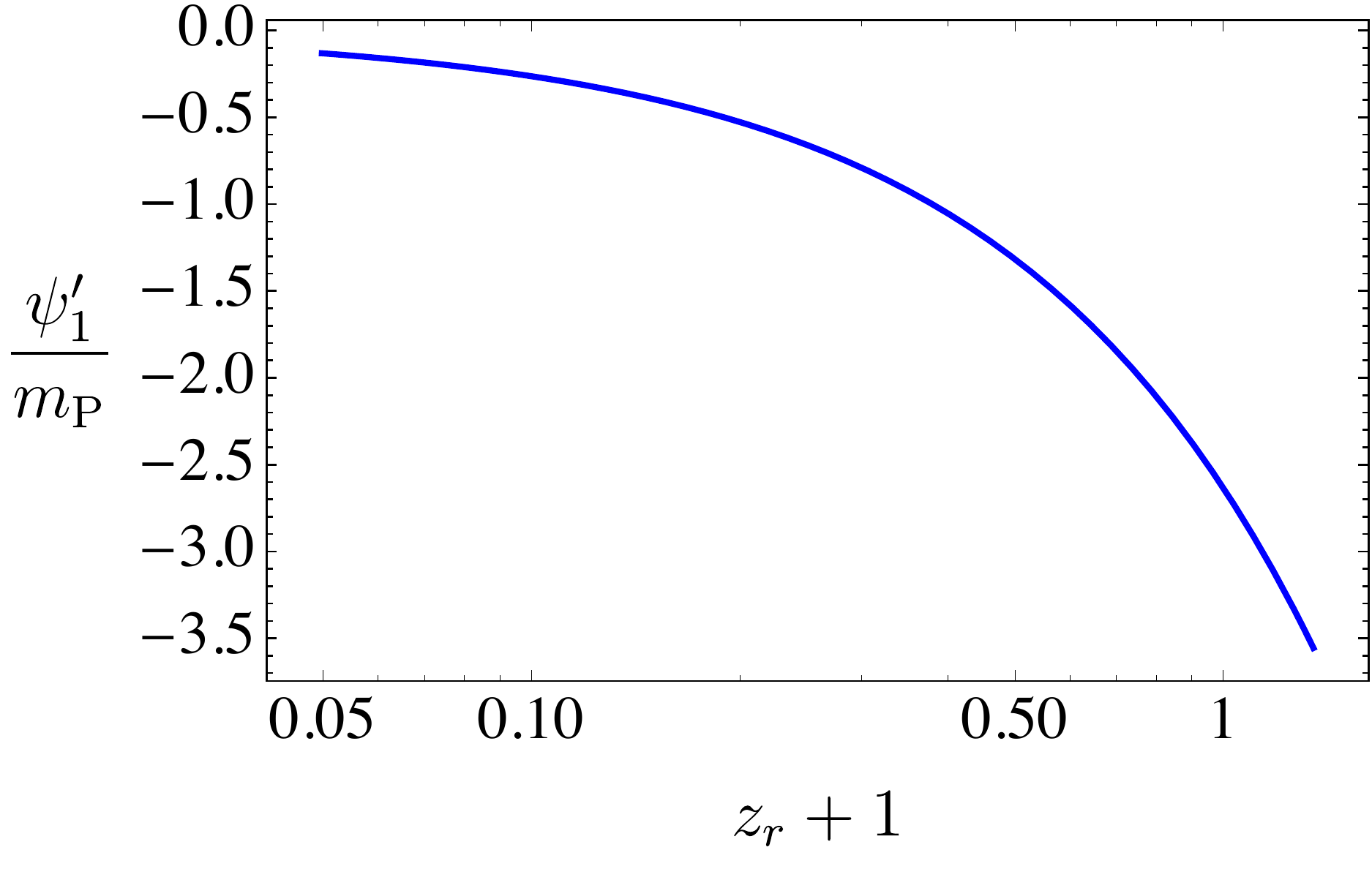}%
}\\
\subfloat[\label{vectorradiation}]{%
  \includegraphics[height=2.5cm,width=.48\linewidth]{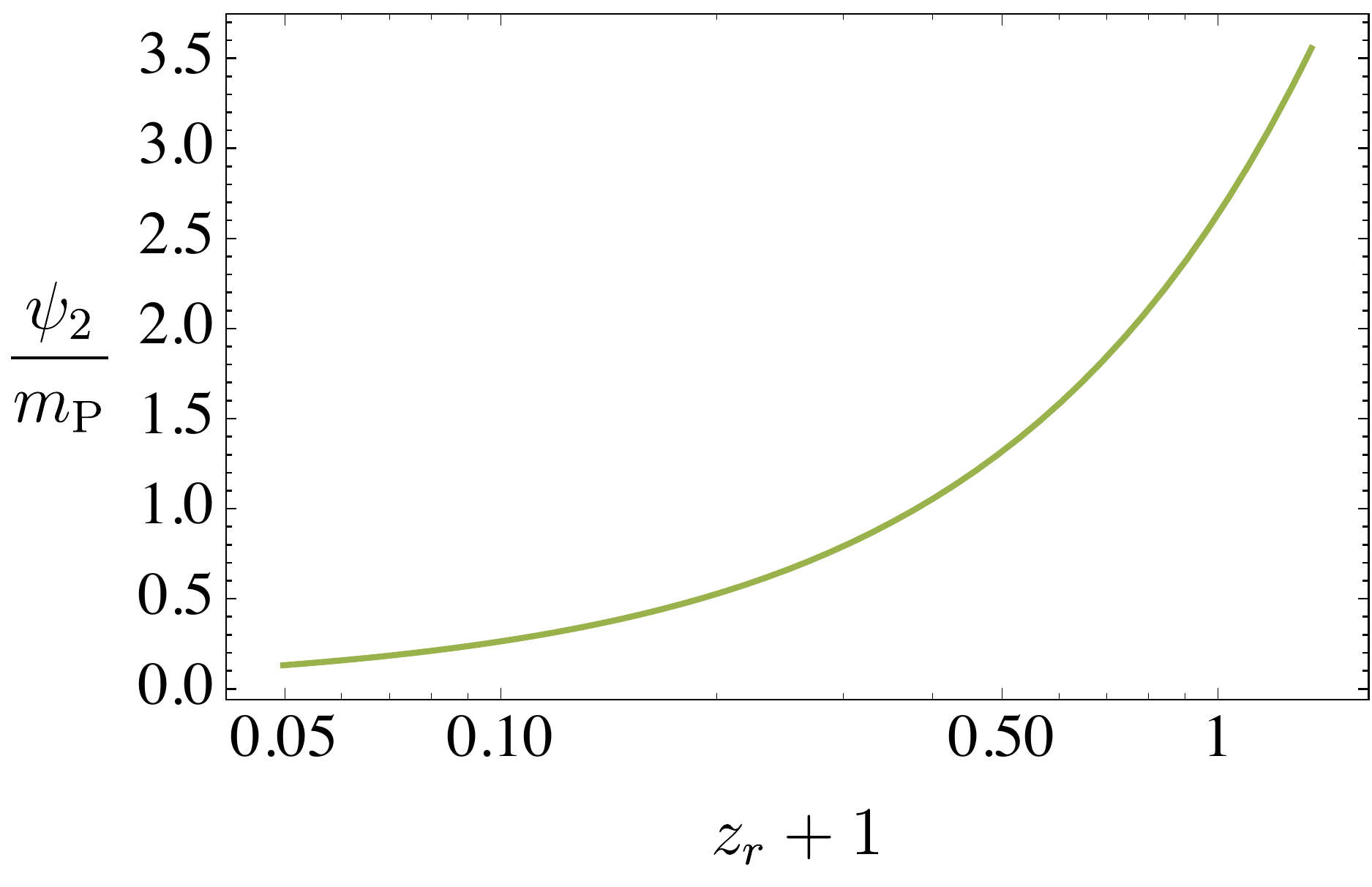}%
}\hfill
\subfloat[\label{vectorspeedradiation}]{%
  \includegraphics[height=2.5cm,width=.48\linewidth]{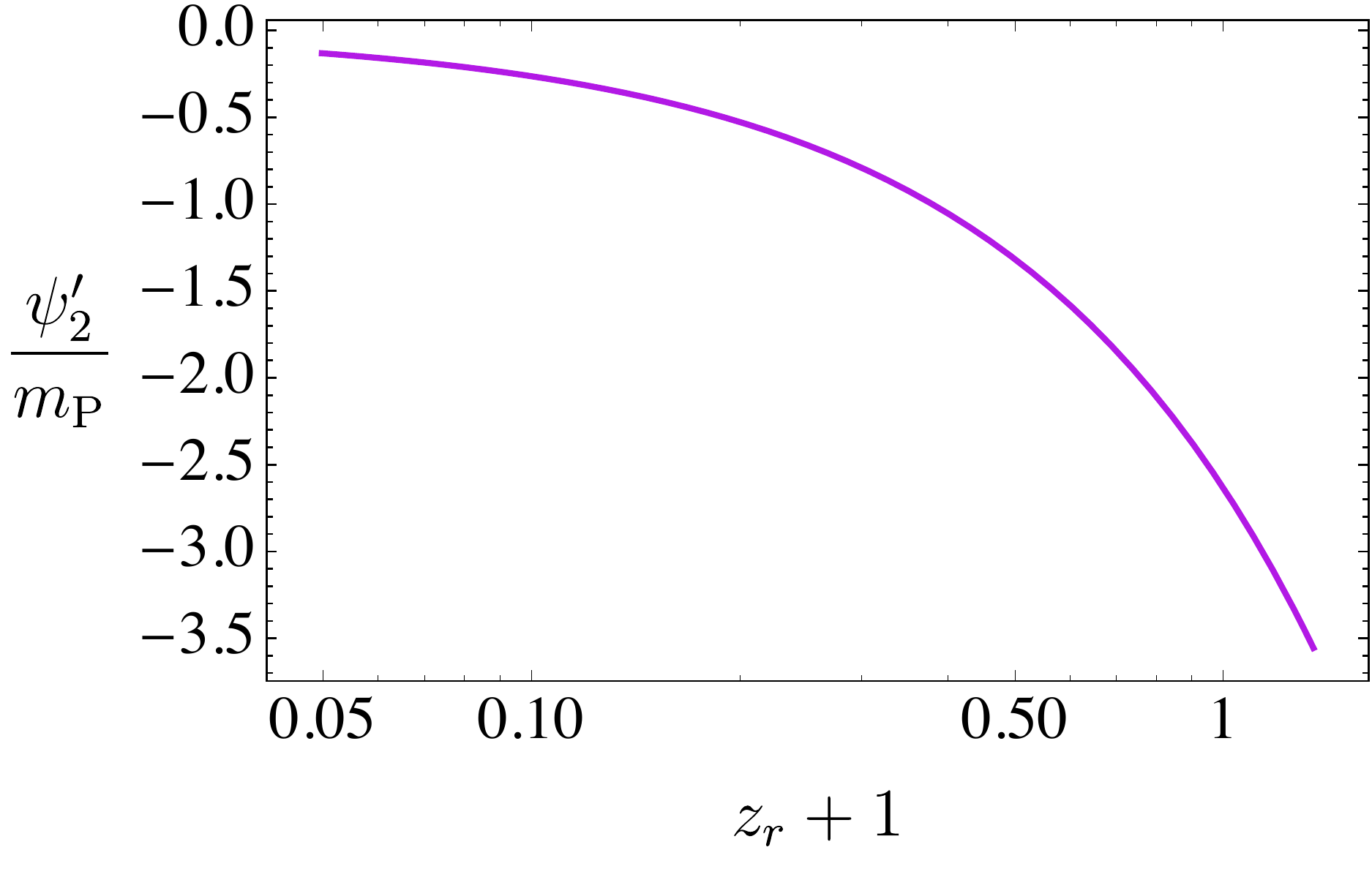}%
}   
\caption{Evolution of the physical fields during the dark energy domination epoch. Plots a) - e) show the speed of the Higgs field, the physical vector fields and their speeds, respectively. The evolution of the Higgs field is not presented since, by this stage, it has almost reached its asymptotic value. The behaviour is in agreement with that obtained via dynamical systems.}
\label{DarkFields}
\end{figure}

This period runs from $z_r \approx 0.3$ onwards in the future since $(\emph{DE-1})$ is the only attractor of the system. The Higgs field reaches a constant value, different to its vacuum value, such that the dominating Higgs potential takes the form of an effective cosmological constant driving the accelerated expansion and the gauge fields almost decay to zero (see Fig. \ref{DarkFields}). When the gauge fields reach their asymptotic values, the interaction between these fields and the Higgs field vanishes and the anisotropic shear has no support. In the particular case fixed by the initial conditions (\ref{Initial conditions}), the shear nowadays\footnote{Here, the subscript $0$ means that the corresponding quantity is evaluated nowadays.} is $\Sigma_0 \approx 2 \times 10^{-13}$, but this value depends on the value $w_i$ as shown below.

\subsubsection{\textbf{Global Evolution}}

\begin{figure}[t]
\includegraphics[width = 0.42\textwidth]{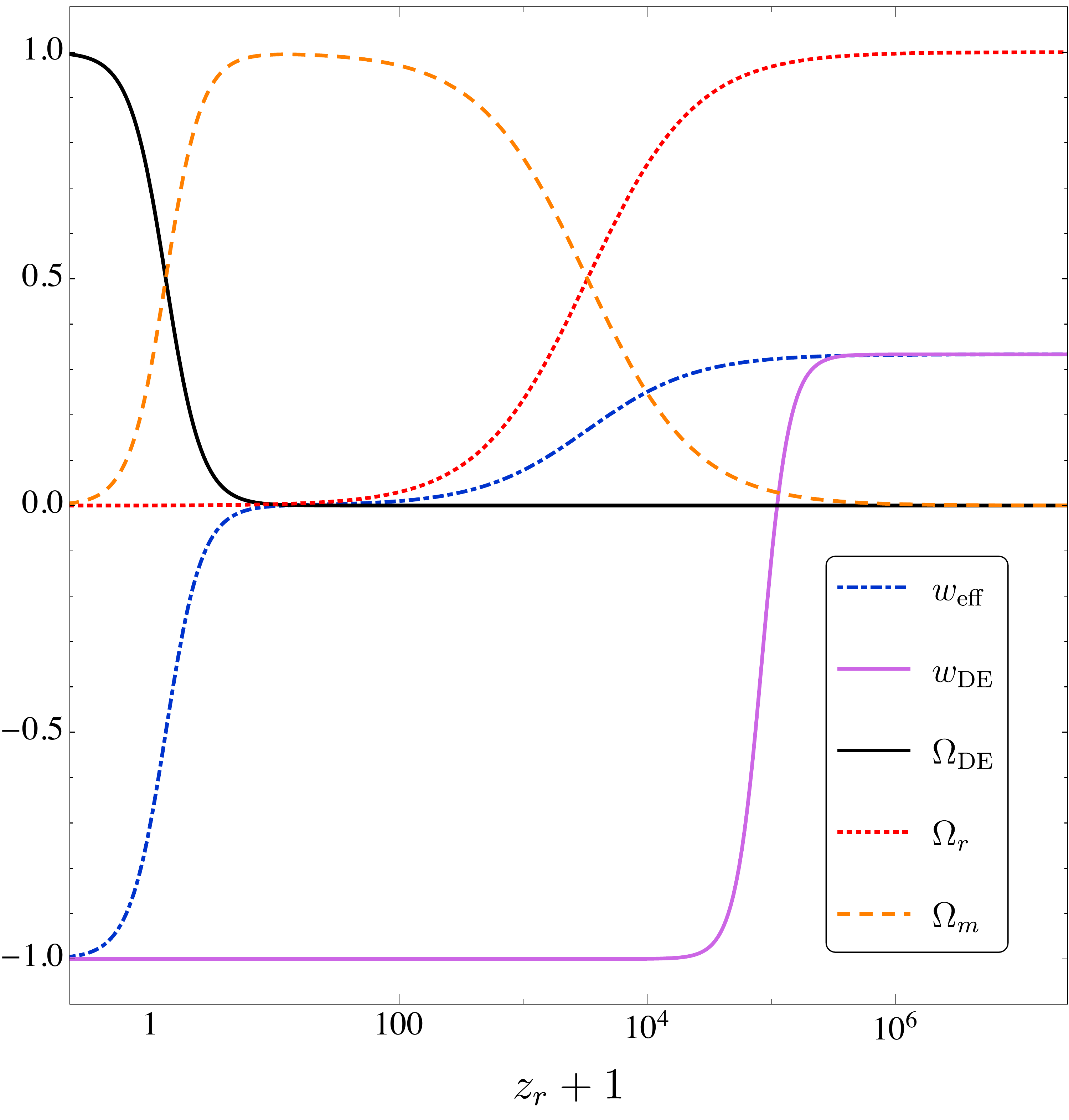}
\caption{Evolution of the density parameters, the effective equation of state and the equation of state of dark energy, during the whole expansion history. The initial conditions (\ref{Initial conditions}) were chosen in the deep radiation era. The Universe transits through a radiation dominance at early times (red dotted line), followed by a matter dominance (light brown dashed line), and ends in the dark energy dominance (black solid line) characterized by $w_{\text{eff}} \simeq -1$ (blue dot-dashed line). The equation of state of dark energy $w_\text{DE}$ (purple solid line) behaves in different ways depending on the epoch.}
\label{Abundances}
\end{figure}

Now, we are going to describe the expansion history of the model in the particular case imposed by the initial conditions (\ref{Initial conditions}). In Fig. \ref{Abundances}, we can see that for large redshifts ($z_r > 10^4$) the Universe is in a radiation dominance epoch, given that $w_{\text{eff}} \simeq 1 / 3$. The equation of state for the dark sector also behaves as a radiation fluid, since $w_\text{DE} \simeq 1 / 3$, meaning that the cosmological trajectory passes near to (some of) the scaling points $(\emph{R-1})$, $(\emph{R-2})$ and $(\emph{R-3})$. At the end of this epoch, $w_{\text{DE}} \simeq - 1$. Around $z_r = 3200$ the Universe experiences the transition between radiation and matter domination, and the matter era begins which is characterized by $w_\text{eff} \simeq 0$. Then, around $z_r = 0.3$ the dark sector becomes dominant and the accelerated expansion of the Universe starts, since $w_{\text{eff}} < - 1 / 3$, quickly getting the value $w_\text{eff} \simeq - 1$. From the end of the radiation epoch, the equation of state of the dark sector stays around the same value $w_\text{DE} \simeq - 1$.

From the dimensionless variables (\ref{variables}), we have been able to obtain an expression for the Hubble parameter:
\begin{equation}
\frac{H}{m_\text{P}} = \frac{\sqrt{3} \, g}{s^2 \, \xi^2} \,,
\end{equation}
where using $H_0 \sim 10^{-61} m_\text{P}$ and the values of $s$ and $\xi$ evaluated today, we have got for the SU(2) coupling parameter $g \sim10^{-88}$. Then, from the parameter $\alpha \equiv \sqrt{2 \lambda / g^2}$ we have got $\lambda \sim 10^{-176}$. We want to stress that these values for $g$ and $\lambda$ are just figures since their actual values depend on the chosen initial conditions; however, they give us a notion of how small they must be.

\subsubsection{\textbf{Anisotropic Shear Today}}

\begin{figure}[t!]
\centering
\includegraphics[width = 0.45\textwidth]{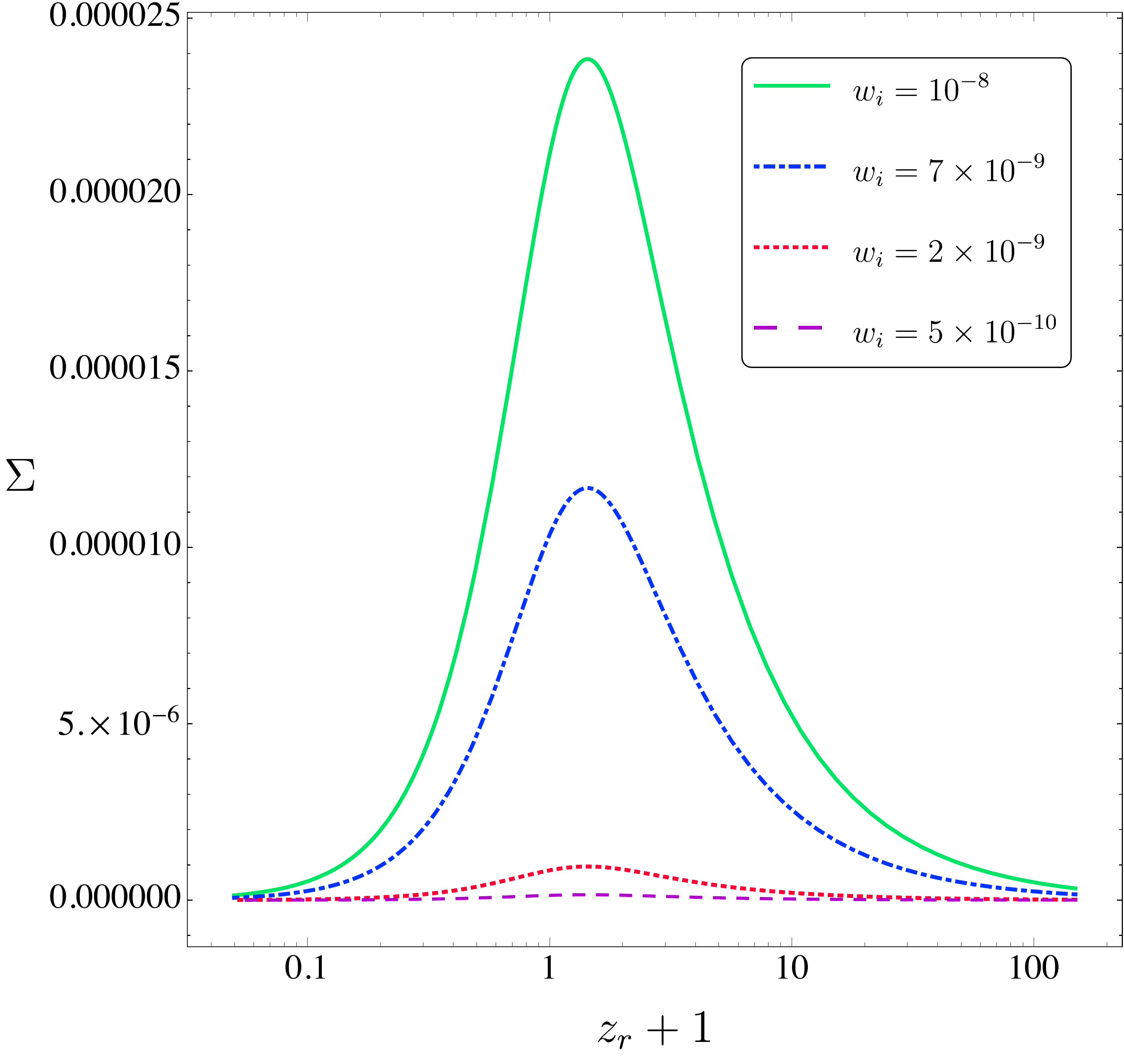}
\caption{
Evolution of the shear $\Sigma$ around $z_r = 0$ for different values of the interaction variable $w$. The shear vanishes in the future in all the cases; however, there are initial conditions yielding to observable values of anisotropic dark energy.}
\label{Anisotropy}
\end{figure}

Observationally, from data analysis of type Ia supernovae, the value of the present shear is constrained to be $\left| \Sigma_0 \right| \leq {\cal O}(0.001)$ \cite{Campanelli:2010zx, Amirhashchi:2018nxl}. However, it is expected future missions like Euclid \cite{Amendola:2016saw} to impose more restricted bounds on the anisotropic contribution from the dark energy content. In this subsection, we are going to explore the behaviour of the anisotropic shear for different initial conditions, and show that the predicted values by our model around the present time, $z_r = 0$, are well within the present bounds. In order to do so, we have fixed the values for the variables as in (\ref{Initial conditions}), except for the variable $w$ which we have varied in a consistent manner with the Friedmann constraint (\ref{Friedmann constraint}), the general behaviour of the density parameters, and the effective equation of state described in the previous subsection. As seen in Fig. \ref{Anisotropy}, the shear vanishes in the future in all the cases studied, being consistent with the dynamical system analysis. We can also see that around the end of the matter epoch ($z_r + 1 \approx 10$) the shear grows to appreciable values, its contribution being non negligible around today ($z_r + 1 = 1$). For instance, for $w_i = 10^{-8}$ we have got $\Sigma_0 \approx 2.1 \times 10^{-5}$, and for $w_i = 5 \times 10^{-10}$ we have got $\Sigma_0 \approx 1.3 \times 10^{-7}$. Therefore, although the Universe loses its hair in the future, it could have an observable hair today;  similar conclusions have been obtained in \cite{Paliathanasis:2020pax} in the framework of the Einstein-aether scalar field theory. 

\subsubsection{\textbf{Equation of State for Dark Energy}}

\begin{figure}[t!]
\centering
\includegraphics[width = 0.45\textwidth]{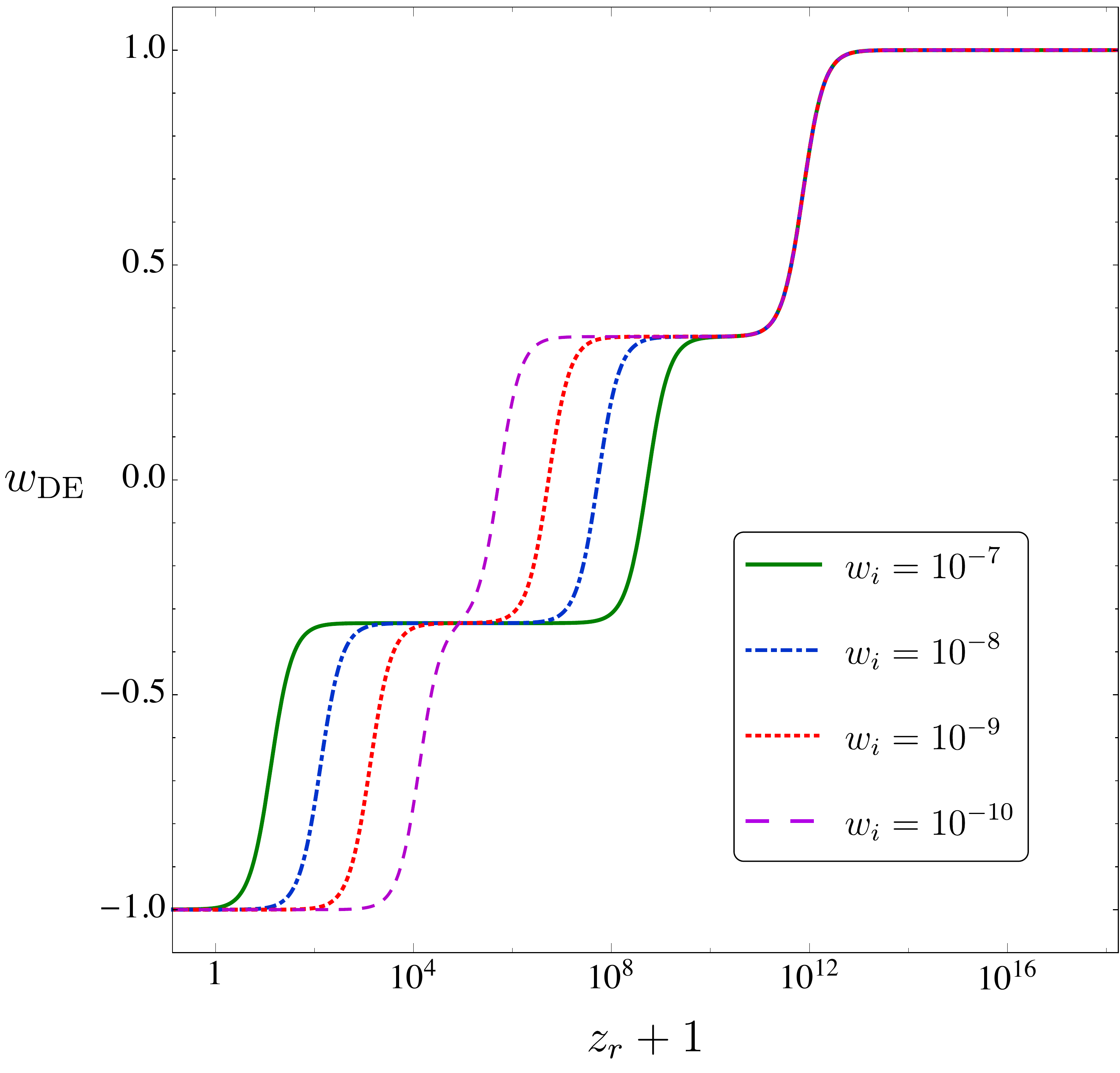}
\caption{Cosmological evolution of the equation of state of dark energy $w_\text{DE}$ for different values of the interaction variable $w$. In all the cases, the final stage is $w_{\text{DE}} \simeq - 1$, and there are also three other stages characterized by $w_{\text{DE}} \simeq 1$, $w_{\text{DE}} \simeq 1 / 3$ and $w_{\text{DE}} \simeq - 1 / 3$.}
\label{DEEoS}
\end{figure}

To end this section, we are going to discuss the cosmological evolution of the equation of state for dark energy $w_\text{DE}$ for different initial conditions. Given that the amount of anisotropic shear is mainly supported by the interaction term encoded in the variable $w$, we are going to explore the behaviour of $w_{\text{DE}}$ with respect to this variable, since we have been interested in the particular trajectories where $\Sigma_0$ is non negligible. Therefore, we fix the initial conditions as in (\ref{Initial conditions}) in the deep radiation era and vary $w_i$ in a consistent way as described in the previous subsection. In Fig. \ref{DEEoS}, we can see that $w_{\text{DE}}$ is characterized by four stages which we are going to describe in the following. 

At very high redshifts, $z_r > 10^{12}$, we have $w_{\text{DE}} \simeq 1$, meaning that the dark sector behaves as a ``stiff fluid" \cite{Zeldovich:1972zz}. During this period, the dark sector is dominated either by the kinetic term of the Higgs field in the variable $z$, phase known as ``kination", or by the shear $\Sigma$. In Appendix \ref{Kination Epoch}, we are going to give some details about this ``kination epoch". After this phase, $w_\text{DE} \simeq 1 / 3$ during the radiation domination, so the dark sector behaves as a ``dark radiation" fluid \cite{Mehrabi:2017xga}. This period corresponds to (some of) the anisotropic scaling points $(\emph{DE-1})$, $(\emph{DE-2})$ and $(\emph{DE-3})$. The length of this period depends on the value of $w$: it increases by decreasing $w$. After this phase, $w_{\text{DE}} \simeq - 1 / 3$, indicating the transition from decelerated to accelerated expansion. This period exists for some values of $w_i$; for instance, in Fig. \ref{Abundances}, this transition phase does not exist, where we have chosen $w_i = 10^{-12}$. This transition phase can be long enough to lasts until the end of the matter era, which has the potential to affect the process of structure formation \cite{Huterer:2013xky}. At the end, in all the cases, the final stage is given by $w_\text{DE} \simeq - 1$, where the symmetry breaking potential dominates and the Higgs field reaches a constant value away from its vacuum value, driving this way an eternal accelerated expansion as an effective cosmological constant.

\section{An Isotropic Possibility: The Higgs Triad} \label{An Isotropic Possibility: The Higgs Triad}

As explained in section \ref{EYMH problems}, a Higgs triplet is inconsistent with the symmetries of the FLRW spacetime. However, there could exist specific arrangements of several fields allowing a homogeneous and isotropic configuration compatible with a FLRW universe. Examples of such arrangements are: a large number of vectors fields pointing at random directions as in vector inflation \cite{Golovnev:2008cf}, the cosmic triad for vector fields charged under some internal symmetry group \cite{Bento:1992wy,ArmendarizPicon:2004pm,Maleknejad:2011jw,Adshead:2012kp,Rodriguez:2017wkg,Gomez:2020sfz}, three inhomogeneous scalar fields as in solid inflation \cite{Endlich:2012pz, ArmendarizPicon:2007nr}, and three U(1) scalar fields as in charged-vector inflation \cite{Firouzjahi:2018wlp}. Based on these works, we propose the ``\emph{Higgs triad}" as a possible configuration consistent with the homogeneity and isotropy of the FLRW metric.

The Higgs triad consists of three SO(3) Higgs fields with the following inner structure
\begin{equation}
\mathcal{H}^I \equiv \begin{pmatrix} 
\Phi \\ 
0 \\
0
\end{pmatrix} \,,\ 
\mathcal{H}^{I I} \equiv \begin{pmatrix} 
0 \\ 
\Phi \\
0
\end{pmatrix} \,,\
\mathcal{H}^{I I} \equiv \begin{pmatrix} 
0 \\ 
0 \\
\Phi
\end{pmatrix} \,,
\end{equation} 
i.e. three Higgs fields with the same magnitude and mutually orthogonal in the manifold associated to the SO(3) group. Each Higgs field comes with its respective gauge vector field, whose dynamics is given by its own Yang-Mills term, and its particular SO(3) gauge coupling. However, in order to keep the isotropy, it is mandatory to consider the same cosmic triad (\ref{ansatz gauge}) for each gauge vector field, and the same group coupling constant $g$. 

Under the conditions mentioned above, the EYMH model matches with a FLRW background. The equations of motion are equivalent to those obtained in section \ref{The Reduced Action}, which were first reported in \cite{Rinaldi:2015iza}. Nonetheless, we want to stress that the original model in \cite{Rinaldi:2015iza} is physically different to the one we are presenting here, since only one Higgs triplet with all its inner components being different is considered there, which yields to the inconsistencies discussed in section \ref{EYMH problems}. 

\section{Conclusions} \label{Conclusions}

In this paper, we have investigated the EYMH model charged under the SO(3) gauge symmetry as a mechanism for the late accelerated expansion. The same model was originally studied in \cite{Rinaldi:2015iza} using the reduced Lagrangian method and a dynamical system analysis. This approach allowed the author of that work to conclude that the model is characterized by a unique accelerated attractor where the potential of the Higgs triplet dominates. However, using the whole set of Einstein equations, we have shown that the equations of motion presented in \cite{Rinaldi:2015iza} are indeed inconsistent with a FLRW background.  This is because the interaction between the gauge components of the Higgs field and the gauge vector field serves as source of momentum density, anisotropic stress or anisotropic shear. Similar inconsistencies were found in \cite{Alvarez:2019ues,Emoto:2002fb,Hosotani:2002nq}. There, the authors studied the EYMH model charged under the SU(2) gauge symmetry, and they found that it is possible to eliminate the inconsistencies by fixing the gauge of the Higgs doublet. However, this is not possible here where the SO(3) gauge group is considered.

In consequence, we have turned our attention to the EYMH model in the SO(3) representation, embedded in a homogeneous but anisotropic axially symmetric Bianchi-I background. We have worked out the respective equations motions, and treated them using a dynamical system approach. We have found that the only attractor of the theory corresponds to an isotropic accelerated expansion where the Higgs field reaches a constant value, different to its vacuum value, such that the dominating Higgs potential behaves as an effective cosmological constant. This result was expected since the interaction term vanishes, which is the support for the spatial shear. However, this does not exclude an observable amount of shear today and, as we have shown above, the model can support non-negligible anisotropies today (for instance, $\Sigma_0 \sim 10^{-5}$) depending on the initial values (in the deep radiation epoch) of the expansion variable $w$ associated to the interaction term. We remark that the values predicted by our model are in agreement with several observational bounds \cite{Campanelli:2010zx, Akarsu:2019pwn,Amirhashchi:2018nxl,Saadeh:2016sak}. Therefore, we conclude that although the Universe loses most of its hair, a small amount could be observable today. We have also found that the equation of state of dark energy $w_\text{DE}$ transits through several characteristic stages depending on the value of $w$. For some values of $w$, during the matter epoch ($z_r < 50$) $w_\text{DE} \neq -1$, suggesting that the process of structure formation might be affected \cite{Huterer:2013xky}. 

\section*{Acknowledgments} 

This work was supported by the following grants: Colciencias-Deutscher Akademischer Austauschdienst Grant No. 110278258747 RC-774-2017, Vicerrector\'{\i}a de Ciencia, Tecnolog\'{\i}a, e Innovaci\'on - Universidad Antonio Nari\~no Grant No. 2019248, Direcci\'on de Investigaci\'on y Extensi\'on de la Facultad de Ciencias - Universidad Industrial de Santander Grant No. 2460, Centro de Investigaciones - Universidad Santo Tom\'as de Aquino Grant No. 1952392, and Vicerrector\'ia de Investigaciones - Universidad del Valle Grant No. 71220. Y.R. wants to acknowledge Massimiliano Rinaldi for useful discussions related to this paper.

\appendix

\section{Kination Epoch} \label{Kination Epoch}

In this appendix, we are going to present another fixed manifold which is irrelevant to the standard radiation, matter and dark energy dominated epochs.

\begin{itemize}
\item Anisotropic Kination $(\emph{K-1})$:
\end{itemize}
\begin{equation*}
x = 0 \,,\ y = 0 \,,\ \xi = 0 \,,\ p = 0 \,,\ s = 0 \,,
\end{equation*}

\begin{equation}
z = \sqrt{1 - \Sigma^2} \,,\ w = 0 \,,\ v = 0 \,,\ \Omega_m = 0 \,.
\end{equation}
This is a non-hyperbolic manifold where the energy budget can be dominated either by the kinetic term of the Higgs field, corresponding to $z = 1$ and known as ``\emph{kination}", or by the spatial shear, corresponding to $\Sigma = 1$. We are going to consider the first option since $\Sigma \approx 0$ is required by observations. In this manifold, we have $w_\text{DE} = 1$ such that the dark sector behaves as a ``stiff fluid" \cite{Zeldovich:1972zz}, and the energy density decays as $\rho_\text{DE} \propto a^{-6}$, meaning that this stiff dominated epoch is prior to the radiation period. These kination solutions are common in quintessence models where the kinetic term of the scalar field has the chance to dominate \cite{Amendola:2015ksp}, and several investigations have pointed out that this period can be useful in the study of the reheating process \cite{Ferreira:1997hj,Pallis:2005bb,Dimopoulos:2018wfg}. However, we are not going to pursue the analysis of this fixed manifold any further.

\bibliography{Bibli.bib} 

\begin{thebibliography}{62}%
\makeatletter
\providecommand \@ifxundefined [1]{%
 \@ifx{#1\undefined}
}%
\providecommand \@ifnum [1]{%
 \ifnum #1\expandafter \@firstoftwo
 \else \expandafter \@secondoftwo
 \fi
}%
\providecommand \@ifx [1]{%
 \ifx #1\expandafter \@firstoftwo
 \else \expandafter \@secondoftwo
 \fi
}%
\providecommand \natexlab [1]{#1}%
\providecommand \enquote  [1]{``#1''}%
\providecommand \bibnamefont  [1]{#1}%
\providecommand \bibfnamefont [1]{#1}%
\providecommand \citenamefont [1]{#1}%
\providecommand \href@noop [0]{\@secondoftwo}%
\providecommand \href [0]{\begingroup \@sanitize@url \@href}%
\providecommand \@href[1]{\@@startlink{#1}\@@href}%
\providecommand \@@href[1]{\endgroup#1\@@endlink}%
\providecommand \@sanitize@url [0]{\catcode `\\12\catcode `\$12\catcode
  `\&12\catcode `\#12\catcode `\^12\catcode `\_12\catcode `\%12\relax}%
\providecommand \@@startlink[1]{}%
\providecommand \@@endlink[0]{}%
\providecommand \url  [0]{\begingroup\@sanitize@url \@url }%
\providecommand \@url [1]{\endgroup\@href {#1}{\urlprefix }}%
\providecommand \urlprefix  [0]{URL }%
\providecommand \Eprint [0]{\href }%
\providecommand \doibase [0]{http://dx.doi.org/}%
\providecommand \selectlanguage [0]{\@gobble}%
\providecommand \bibinfo  [0]{\@secondoftwo}%
\providecommand \bibfield  [0]{\@secondoftwo}%
\providecommand \translation [1]{[#1]}%
\providecommand \BibitemOpen [0]{}%
\providecommand \bibitemStop [0]{}%
\providecommand \bibitemNoStop [0]{.\EOS\space}%
\providecommand \EOS [0]{\spacefactor3000\relax}%
\providecommand \BibitemShut  [1]{\csname bibitem#1\endcsname}%
\let\auto@bib@innerbib\@empty
\bibitem [{\citenamefont {Riess}\ \emph {et~al.}(1998)\citenamefont {Riess}
  \emph {et~al.}}]{Riess:1998cb}%
  \BibitemOpen
  \bibfield  {author} {\bibinfo {author} {\bibfnamefont {A.~G.}\ \bibnamefont
  {Riess}} \emph {et~al.} (\bibinfo {collaboration} {Supernova Search Team}),\
  }\bibfield  {title} {\enquote {\bibinfo {title} {{Observational evidence from
  supernovae for an accelerating universe and a cosmological constant}},}\
  }\href {\doibase 10.1086/300499} {\bibfield  {journal} {\bibinfo  {journal}
  {Astron. J.}\ }\textbf {\bibinfo {volume} {116}},\ \bibinfo {pages}
  {1009--1038} (\bibinfo {year} {1998})},\ \Eprint
  {http://arxiv.org/abs/astro-ph/9805201} {arXiv:astro-ph/9805201} \BibitemShut
  {NoStop}%
\bibitem [{\citenamefont {Perlmutter}\ \emph {et~al.}(1999)\citenamefont
  {Perlmutter} \emph {et~al.}}]{Perlmutter:1998np}%
  \BibitemOpen
  \bibfield  {author} {\bibinfo {author} {\bibfnamefont {S.}~\bibnamefont
  {Perlmutter}} \emph {et~al.} (\bibinfo {collaboration} {Supernova Cosmology
  Project}),\ }\bibfield  {title} {\enquote {\bibinfo {title} {{Measurements of
  $\Omega$ and $\Lambda$ from 42 high redshift supernovae}},}\ }\href {\doibase
  10.1086/307221} {\bibfield  {journal} {\bibinfo  {journal} {Astrophys. J.}\
  }\textbf {\bibinfo {volume} {517}},\ \bibinfo {pages} {565--586} (\bibinfo
  {year} {1999})},\ \Eprint {http://arxiv.org/abs/astro-ph/9812133}
  {arXiv:astro-ph/9812133} \BibitemShut {NoStop}%
\bibitem [{\citenamefont {Aghanim}\ \emph {et~al.}(2020)\citenamefont {Aghanim}
  \emph {et~al.}}]{Aghanim:2018eyx}%
  \BibitemOpen
  \bibfield  {author} {\bibinfo {author} {\bibfnamefont {N.}~\bibnamefont
  {Aghanim}} \emph {et~al.} (\bibinfo {collaboration} {Planck}),\ }\bibfield
  {title} {\enquote {\bibinfo {title} {{Planck 2018 results. VI. Cosmological
  parameters}},}\ }\href {\doibase 10.1051/0004-6361/201833910} {\bibfield
  {journal} {\bibinfo  {journal} {Astron. Astrophys.}\ }\textbf {\bibinfo
  {volume} {641}},\ \bibinfo {pages} {A6} (\bibinfo {year} {2020})},\ \Eprint
  {http://arxiv.org/abs/1807.06209} {arXiv:1807.06209 [astro-ph.CO]}
  \BibitemShut {NoStop}%
\bibitem [{\citenamefont {Troxel}\ \emph {et~al.}(2018)\citenamefont {Troxel}
  \emph {et~al.}}]{Troxel:2017xyo}%
  \BibitemOpen
  \bibfield  {author} {\bibinfo {author} {\bibfnamefont {M.~A.}\ \bibnamefont
  {Troxel}} \emph {et~al.} (\bibinfo {collaboration} {DES}),\ }\bibfield
  {title} {\enquote {\bibinfo {title} {{Dark Energy Survey Year 1 results:
  Cosmological constraints from cosmic shear}},}\ }\href {\doibase
  10.1103/PhysRevD.98.043528} {\bibfield  {journal} {\bibinfo  {journal} {Phys.
  Rev. D}\ }\textbf {\bibinfo {volume} {98}},\ \bibinfo {pages} {043528}
  (\bibinfo {year} {2018})},\ \Eprint {http://arxiv.org/abs/1708.01538}
  {arXiv:1708.01538 [astro-ph.CO]} \BibitemShut {NoStop}%
\bibitem [{\citenamefont {Abbott}\ \emph
  {et~al.}(2019{\natexlab{a}})\citenamefont {Abbott} \emph
  {et~al.}}]{Abbott:2018xao}%
  \BibitemOpen
  \bibfield  {author} {\bibinfo {author} {\bibfnamefont {T.~M.~C.}\
  \bibnamefont {Abbott}} \emph {et~al.} (\bibinfo {collaboration} {DES}),\
  }\bibfield  {title} {\enquote {\bibinfo {title} {{Dark Energy Survey Year 1
  Results: Constraints on Extended Cosmological Models from Galaxy Clustering
  and Weak Lensing}},}\ }\href {\doibase 10.1103/PhysRevD.99.123505} {\bibfield
   {journal} {\bibinfo  {journal} {Phys. Rev. D}\ }\textbf {\bibinfo {volume}
  {99}},\ \bibinfo {pages} {123505} (\bibinfo {year} {2019}{\natexlab{a}})},\
  \Eprint {http://arxiv.org/abs/1810.02499} {arXiv:1810.02499 [astro-ph.CO]}
  \BibitemShut {NoStop}%
\bibitem [{\citenamefont {Sahni}\ \emph {et~al.}(2014)\citenamefont {Sahni},
  \citenamefont {Shafieloo},\ and\ \citenamefont
  {Starobinsky}}]{Sahni:2014ooa}%
  \BibitemOpen
  \bibfield  {author} {\bibinfo {author} {\bibfnamefont {V.}~\bibnamefont
  {Sahni}}, \bibinfo {author} {\bibfnamefont {A.}~\bibnamefont {Shafieloo}}, \
  and\ \bibinfo {author} {\bibfnamefont {A.~A.}\ \bibnamefont {Starobinsky}},\
  }\bibfield  {title} {\enquote {\bibinfo {title} {{Model independent evidence
  for dark energy evolution from Baryon Acoustic Oscillations}},}\ }\href
  {\doibase 10.1088/2041-8205/793/2/L40} {\bibfield  {journal} {\bibinfo
  {journal} {Astrophys. J.}\ }\textbf {\bibinfo {volume} {793}},\ \bibinfo
  {pages} {L40} (\bibinfo {year} {2014})},\ \Eprint
  {http://arxiv.org/abs/1406.2209} {arXiv:1406.2209 [astro-ph.CO]} \BibitemShut
  {NoStop}%
\bibitem [{\citenamefont {Weinberg}(1989)}]{Weinberg:1988cp}%
  \BibitemOpen
  \bibfield  {author} {\bibinfo {author} {\bibfnamefont {S.}~\bibnamefont
  {Weinberg}},\ }\bibfield  {title} {\enquote {\bibinfo {title} {{The
  Cosmological Constant Problem}},}\ }\href {\doibase 10.1103/RevModPhys.61.1}
  {\bibfield  {journal} {\bibinfo  {journal} {Rev. Mod. Phys.}\ }\textbf
  {\bibinfo {volume} {61}},\ \bibinfo {pages} {1--23} (\bibinfo {year}
  {1989})}\BibitemShut {NoStop}%
\bibitem [{\citenamefont {Di~Valentino}\ \emph {et~al.}(2019)\citenamefont
  {Di~Valentino}, \citenamefont {Melchiorri},\ and\ \citenamefont
  {Silk}}]{DiValentino:2019qzk}%
  \BibitemOpen
  \bibfield  {author} {\bibinfo {author} {\bibfnamefont {E.}~\bibnamefont
  {Di~Valentino}}, \bibinfo {author} {\bibfnamefont {A.}~\bibnamefont
  {Melchiorri}}, \ and\ \bibinfo {author} {\bibfnamefont {J.}~\bibnamefont
  {Silk}},\ }\bibfield  {title} {\enquote {\bibinfo {title} {{Planck evidence
  for a closed Universe and a possible crisis for cosmology}},}\ }\href
  {\doibase 10.1038/s41550-019-0906-9} {\bibfield  {journal} {\bibinfo
  {journal} {Nature Astron.}\ }\textbf {\bibinfo {volume} {4}},\ \bibinfo
  {pages} {196--203} (\bibinfo {year} {2019})},\ \Eprint
  {http://arxiv.org/abs/1911.02087} {arXiv:1911.02087 [astro-ph.CO]}
  \BibitemShut {NoStop}%
\bibitem [{\citenamefont {Verde}\ \emph {et~al.}(2019)\citenamefont {Verde},
  \citenamefont {Treu},\ and\ \citenamefont {Riess}}]{Verde:2019ivm}%
  \BibitemOpen
  \bibfield  {author} {\bibinfo {author} {\bibfnamefont {L.}~\bibnamefont
  {Verde}}, \bibinfo {author} {\bibfnamefont {T.}~\bibnamefont {Treu}}, \ and\
  \bibinfo {author} {\bibfnamefont {A.~G.}\ \bibnamefont {Riess}},\ }\bibfield
  {title} {\enquote {\bibinfo {title} {{Tensions between the Early and the Late
  Universe}},}\ }\href {\doibase 10.1038/s41550-019-0902-0} {\bibfield
  {journal} {\bibinfo  {journal} {Nature Astron.}\ }\textbf {\bibinfo {volume}
  {3}},\ \bibinfo {pages} {891--895} (\bibinfo {year} {2019})},\ \Eprint
  {http://arxiv.org/abs/1907.10625} {arXiv:1907.10625 [astro-ph.CO]}
  \BibitemShut {NoStop}%
\bibitem [{\citenamefont {Handley}(2019)}]{Handley:2019tkm}%
  \BibitemOpen
  \bibfield  {author} {\bibinfo {author} {\bibfnamefont {W.}~\bibnamefont
  {Handley}},\ }\bibfield  {title} {\enquote {\bibinfo {title} {{Curvature
  tension: evidence for a closed universe}},}\ }\href@noop {} {\  (\bibinfo
  {year} {2019})},\ \Eprint {http://arxiv.org/abs/1908.09139} {arXiv:1908.09139
  [astro-ph.CO]} \BibitemShut {NoStop}%
\bibitem [{\citenamefont {Guo}\ \emph {et~al.}(2019)\citenamefont {Guo},
  \citenamefont {Zhang},\ and\ \citenamefont {Zhang}}]{Guo:2018ans}%
  \BibitemOpen
  \bibfield  {author} {\bibinfo {author} {\bibfnamefont {R.-y.}\ \bibnamefont
  {Guo}}, \bibinfo {author} {\bibfnamefont {J.-f.}\ \bibnamefont {Zhang}}, \
  and\ \bibinfo {author} {\bibfnamefont {X.}~\bibnamefont {Zhang}},\ }\bibfield
   {title} {\enquote {\bibinfo {title} {{Can the $H_0$ tension be resolved in
  extensions to $\Lambda$CDM cosmology?}}}\ }\href {\doibase
  10.1088/1475-7516/2019/02/054} {\bibfield  {journal} {\bibinfo  {journal}
  {JCAP}\ }\textbf {\bibinfo {volume} {1902}},\ \bibinfo {pages} {054}
  (\bibinfo {year} {2019})},\ \Eprint {http://arxiv.org/abs/1809.02340}
  {arXiv:1809.02340 [astro-ph.CO]} \BibitemShut {NoStop}%
\bibitem [{\citenamefont {Collett}\ \emph {et~al.}(2018)\citenamefont {Collett}
  \emph {et~al.}}]{Collett:2018gpf}%
  \BibitemOpen
  \bibfield  {author} {\bibinfo {author} {\bibfnamefont {T.~E.}\ \bibnamefont
  {Collett}} \emph {et~al.},\ }\bibfield  {title} {\enquote {\bibinfo {title}
  {{A precise extragalactic test of General Relativity}},}\ }\href {\doibase
  10.1126/science.aao2469} {\bibfield  {journal} {\bibinfo  {journal}
  {Science}\ }\textbf {\bibinfo {volume} {360}},\ \bibinfo {pages} {1342}
  (\bibinfo {year} {2018})},\ \Eprint {http://arxiv.org/abs/1806.08300}
  {arXiv:1806.08300 [astro-ph.CO]} \BibitemShut {NoStop}%
\bibitem [{\citenamefont {Ezquiaga}\ and\ \citenamefont
  {Zumalac{\'a}rregui}(2018)}]{Ezquiaga:2018btd}%
  \BibitemOpen
  \bibfield  {author} {\bibinfo {author} {\bibfnamefont {J.~M.}\ \bibnamefont
  {Ezquiaga}}\ and\ \bibinfo {author} {\bibfnamefont {M.}~\bibnamefont
  {Zumalac{\'a}rregui}},\ }\bibfield  {title} {\enquote {\bibinfo {title}
  {{Dark Energy in light of Multi-Messenger Gravitational-Wave astronomy}},}\
  }\href {\doibase 10.3389/fspas.2018.00044} {\bibfield  {journal} {\bibinfo
  {journal} {Front. Astron. Space Sci.}\ }\textbf {\bibinfo {volume} {5}},\
  \bibinfo {pages} {44} (\bibinfo {year} {2018})},\ \Eprint
  {http://arxiv.org/abs/1807.09241} {arXiv:1807.09241 [astro-ph.CO]}
  \BibitemShut {NoStop}%
\bibitem [{\citenamefont {He}\ \emph {et~al.}(2018)\citenamefont {He},
  \citenamefont {Guzzo}, \citenamefont {Li},\ and\ \citenamefont
  {Baugh}}]{He:2018oai}%
  \BibitemOpen
  \bibfield  {author} {\bibinfo {author} {\bibfnamefont {J.-h.}\ \bibnamefont
  {He}}, \bibinfo {author} {\bibfnamefont {L.}~\bibnamefont {Guzzo}}, \bibinfo
  {author} {\bibfnamefont {B.}~\bibnamefont {Li}}, \ and\ \bibinfo {author}
  {\bibfnamefont {C.~M.}\ \bibnamefont {Baugh}},\ }\bibfield  {title} {\enquote
  {\bibinfo {title} {{No evidence for modifications of gravity from galaxy
  motions on cosmological scales}},}\ }\href {\doibase
  10.1038/s41550-018-0573-2} {\bibfield  {journal} {\bibinfo  {journal} {Nature
  Astron.}\ }\textbf {\bibinfo {volume} {2}},\ \bibinfo {pages} {967--972}
  (\bibinfo {year} {2018})},\ \Eprint {http://arxiv.org/abs/1809.09019}
  {arXiv:1809.09019 [astro-ph.CO]} \BibitemShut {NoStop}%
\bibitem [{\citenamefont {Do}\ \emph {et~al.}(2019)\citenamefont {Do} \emph
  {et~al.}}]{Do:2019txf}%
  \BibitemOpen
  \bibfield  {author} {\bibinfo {author} {\bibfnamefont {T.}~\bibnamefont {Do}}
  \emph {et~al.},\ }\bibfield  {title} {\enquote {\bibinfo {title}
  {{Relativistic redshift of the star S0-2 orbiting the galactic center
  supermassive black hole}},}\ }\href {\doibase 10.1126/science.aav8137}
  {\bibfield  {journal} {\bibinfo  {journal} {Science}\ }\textbf {\bibinfo
  {volume} {365}},\ \bibinfo {pages} {664--668} (\bibinfo {year} {2019})},\
  \Eprint {http://arxiv.org/abs/1907.10731} {arXiv:1907.10731 [astro-ph.GA]}
  \BibitemShut {NoStop}%
\bibitem [{\citenamefont {Abbott}\ \emph
  {et~al.}(2019{\natexlab{b}})\citenamefont {Abbott} \emph
  {et~al.}}]{Abbott:2018lct}%
  \BibitemOpen
  \bibfield  {author} {\bibinfo {author} {\bibfnamefont {B.~P.}\ \bibnamefont
  {Abbott}} \emph {et~al.} (\bibinfo {collaboration} {LIGO Scientific,
  Virgo}),\ }\bibfield  {title} {\enquote {\bibinfo {title} {{Tests of General
  Relativity with GW170817}},}\ }\href {\doibase
  10.1103/PhysRevLett.123.011102} {\bibfield  {journal} {\bibinfo  {journal}
  {Phys. Rev. Lett.}\ }\textbf {\bibinfo {volume} {123}},\ \bibinfo {pages}
  {011102} (\bibinfo {year} {2019}{\natexlab{b}})},\ \Eprint
  {http://arxiv.org/abs/1811.00364} {arXiv:1811.00364 [gr-qc]} \BibitemShut
  {NoStop}%
\bibitem [{\citenamefont {Akiyama}\ \emph {et~al.}(2019)\citenamefont {Akiyama}
  \emph {et~al.}}]{Akiyama:2019cqa}%
  \BibitemOpen
  \bibfield  {author} {\bibinfo {author} {\bibfnamefont {K.}~\bibnamefont
  {Akiyama}} \emph {et~al.} (\bibinfo {collaboration} {Event Horizon
  Telescope}),\ }\bibfield  {title} {\enquote {\bibinfo {title} {{First M87
  Event Horizon Telescope Results. I. The Shadow of the Supermassive Black
  Hole}},}\ }\href {\doibase 10.3847/2041-8213/ab0ec7} {\bibfield  {journal}
  {\bibinfo  {journal} {Astrophys. J.}\ }\textbf {\bibinfo {volume} {875}},\
  \bibinfo {pages} {L1} (\bibinfo {year} {2019})},\ \Eprint
  {http://arxiv.org/abs/1906.11238} {arXiv:1906.11238 [astro-ph.GA]}
  \BibitemShut {NoStop}%
\bibitem [{\citenamefont {Kane}(2017)}]{Kane:1987gb}%
  \BibitemOpen
  \bibfield  {author} {\bibinfo {author} {\bibfnamefont {G.~L.}\ \bibnamefont
  {Kane}},\ }\href
  {http://www.cambridge.org/academic/subjects/physics/particle-physics-and-nuclear-physics/modern-elementary-particle-physics-explaining-and-extending-standard-model-2nd-edition?format=AR&isbn=9781316730805}
  {\emph {\bibinfo {title} {{Modern Elementary Particle Physics}}}}\ (\bibinfo
  {publisher} {Cambridge University Press},\ \bibinfo {year}
  {2017})\BibitemShut {NoStop}%
\bibitem [{\citenamefont {Rinaldi}(2015{\natexlab{a}})}]{Rinaldi:2014yta}%
  \BibitemOpen
  \bibfield  {author} {\bibinfo {author} {\bibfnamefont {M.}~\bibnamefont
  {Rinaldi}},\ }\bibfield  {title} {\enquote {\bibinfo {title} {{Higgs Dark
  Energy}},}\ }\href {\doibase 10.1088/0264-9381/32/4/045002} {\bibfield
  {journal} {\bibinfo  {journal} {Class. Quant. Grav.}\ }\textbf {\bibinfo
  {volume} {32}},\ \bibinfo {pages} {045002} (\bibinfo {year}
  {2015}{\natexlab{a}})},\ \Eprint {http://arxiv.org/abs/1404.0532}
  {arXiv:1404.0532 [astro-ph.CO]} \BibitemShut {NoStop}%
\bibitem [{\citenamefont {Boyle}\ \emph {et~al.}(2002)\citenamefont {Boyle},
  \citenamefont {Caldwell},\ and\ \citenamefont {Kamionkowski}}]{Boyle:2001du}%
  \BibitemOpen
  \bibfield  {author} {\bibinfo {author} {\bibfnamefont {L.~A.}\ \bibnamefont
  {Boyle}}, \bibinfo {author} {\bibfnamefont {R.~R.}\ \bibnamefont {Caldwell}},
  \ and\ \bibinfo {author} {\bibfnamefont {M.}~\bibnamefont {Kamionkowski}},\
  }\bibfield  {title} {\enquote {\bibinfo {title} {{Spintessence! New models
  for dark matter and dark energy}},}\ }\href {\doibase
  10.1016/S0370-2693(02)02590-X} {\bibfield  {journal} {\bibinfo  {journal}
  {Phys. Lett. B}\ }\textbf {\bibinfo {volume} {545}},\ \bibinfo {pages}
  {17--22} (\bibinfo {year} {2002})},\ \Eprint
  {http://arxiv.org/abs/astro-ph/0105318} {arXiv:astro-ph/0105318} \BibitemShut
  {NoStop}%
\bibitem [{\citenamefont {{\'A}lvarez}\ \emph {et~al.}(2019)\citenamefont
  {{\'A}lvarez}, \citenamefont {Orjuela-Quintana}, \citenamefont
  {Rodr\'{\i}guez},\ and\ \citenamefont {Valenzuela-Toledo}}]{Alvarez:2019ues}%
  \BibitemOpen
  \bibfield  {author} {\bibinfo {author} {\bibfnamefont {M.}~\bibnamefont
  {{\'A}lvarez}}, \bibinfo {author} {\bibfnamefont {J.~B.}\ \bibnamefont
  {Orjuela-Quintana}}, \bibinfo {author} {\bibfnamefont {Y.}~\bibnamefont
  {Rodr\'{\i}guez}}, \ and\ \bibinfo {author} {\bibfnamefont {C.~A.}\
  \bibnamefont {Valenzuela-Toledo}},\ }\bibfield  {title} {\enquote {\bibinfo
  {title} {{Einstein Yang--Mills Higgs dark energy revisited}},}\ }\href
  {\doibase 10.1088/1361-6382/ab3775} {\bibfield  {journal} {\bibinfo
  {journal} {Class. Quant. Grav.}\ }\textbf {\bibinfo {volume} {36}},\ \bibinfo
  {pages} {195004} (\bibinfo {year} {2019})},\ \Eprint
  {http://arxiv.org/abs/1901.04624} {arXiv:1901.04624 [gr-qc]} \BibitemShut
  {NoStop}%
\bibitem [{\citenamefont {Emoto}\ \emph {et~al.}(2002)\citenamefont {Emoto},
  \citenamefont {Hosotani},\ and\ \citenamefont {Kubota}}]{Emoto:2002fb}%
  \BibitemOpen
  \bibfield  {author} {\bibinfo {author} {\bibfnamefont {H.}~\bibnamefont
  {Emoto}}, \bibinfo {author} {\bibfnamefont {Y.}~\bibnamefont {Hosotani}}, \
  and\ \bibinfo {author} {\bibfnamefont {T.}~\bibnamefont {Kubota}},\
  }\bibfield  {title} {\enquote {\bibinfo {title} {{Cosmology in the Einstein
  electroweak theory and magnetic fields}},}\ }\href {\doibase
  10.1143/PTP.108.157} {\bibfield  {journal} {\bibinfo  {journal} {Prog. Theor.
  Phys.}\ }\textbf {\bibinfo {volume} {108}},\ \bibinfo {pages} {157--183}
  (\bibinfo {year} {2002})},\ \Eprint {http://arxiv.org/abs/hep-th/0201141}
  {arXiv:hep-th/0201141} \BibitemShut {NoStop}%
\bibitem [{\citenamefont {Hosotani}\ \emph {et~al.}(2003)\citenamefont
  {Hosotani}, \citenamefont {Emoto},\ and\ \citenamefont
  {Kubota}}]{Hosotani:2002nq}%
  \BibitemOpen
  \bibfield  {author} {\bibinfo {author} {\bibfnamefont {Y.}~\bibnamefont
  {Hosotani}}, \bibinfo {author} {\bibfnamefont {H.}~\bibnamefont {Emoto}}, \
  and\ \bibinfo {author} {\bibfnamefont {T.}~\bibnamefont {Kubota}},\
  }\bibfield  {title} {\enquote {\bibinfo {title} {{Cosmic solutions in the
  Einstein-Weinberg-Salam theory and the generation of large electric and
  magnetic fields}},}\ }\href {\doibase 10.1016/S0920-5632(03)90510-X}
  {\bibfield  {journal} {\bibinfo  {journal} {Nucl. Phys. B Proc. Suppl.}\
  }\textbf {\bibinfo {volume} {117}},\ \bibinfo {pages} {139} (\bibinfo {year}
  {2003})},\ \Eprint {http://arxiv.org/abs/hep-ph/0209112}
  {arXiv:hep-ph/0209112} \BibitemShut {NoStop}%
\bibitem [{\citenamefont {Rinaldi}(2015{\natexlab{b}})}]{Rinaldi:2015iza}%
  \BibitemOpen
  \bibfield  {author} {\bibinfo {author} {\bibfnamefont {M.}~\bibnamefont
  {Rinaldi}},\ }\bibfield  {title} {\enquote {\bibinfo {title} {{Dark energy as
  a fixed point of the Einstein Yang-Mills Higgs Equations}},}\ }\href
  {\doibase 10.1088/1475-7516/2015/10/023} {\bibfield  {journal} {\bibinfo
  {journal} {JCAP}\ }\textbf {\bibinfo {volume} {1510}},\ \bibinfo {pages}
  {023} (\bibinfo {year} {2015}{\natexlab{b}})},\ \Eprint
  {http://arxiv.org/abs/1508.04576} {arXiv:1508.04576 [gr-qc]} \BibitemShut
  {NoStop}%
\bibitem [{\citenamefont {Koivisto}\ and\ \citenamefont
  {Mota}(2008)}]{Koivisto:2008ig}%
  \BibitemOpen
  \bibfield  {author} {\bibinfo {author} {\bibfnamefont {T.}~\bibnamefont
  {Koivisto}}\ and\ \bibinfo {author} {\bibfnamefont {D.~F.}\ \bibnamefont
  {Mota}},\ }\bibfield  {title} {\enquote {\bibinfo {title} {{Anisotropic Dark
  Energy: Dynamics of Background and Perturbations}},}\ }\href {\doibase
  10.1088/1475-7516/2008/06/018} {\bibfield  {journal} {\bibinfo  {journal}
  {JCAP}\ }\textbf {\bibinfo {volume} {0806}},\ \bibinfo {pages} {018}
  (\bibinfo {year} {2008})},\ \Eprint {http://arxiv.org/abs/0801.3676}
  {arXiv:0801.3676 [astro-ph]} \BibitemShut {NoStop}%
\bibitem [{\citenamefont {Adshead}\ and\ \citenamefont
  {Liu}(2018)}]{Adshead:2018emn}%
  \BibitemOpen
  \bibfield  {author} {\bibinfo {author} {\bibfnamefont {P.}~\bibnamefont
  {Adshead}}\ and\ \bibinfo {author} {\bibfnamefont {A.}~\bibnamefont {Liu}},\
  }\bibfield  {title} {\enquote {\bibinfo {title} {{Anisotropic Massive
  Gauge-flation}},}\ }\href {\doibase 10.1088/1475-7516/2018/07/052} {\bibfield
   {journal} {\bibinfo  {journal} {JCAP}\ }\textbf {\bibinfo {volume} {1807}},\
  \bibinfo {pages} {052} (\bibinfo {year} {2018})},\ \Eprint
  {http://arxiv.org/abs/1803.07168} {arXiv:1803.07168 [astro-ph.CO]}
  \BibitemShut {NoStop}%
\bibitem [{\citenamefont {Beltr{\'a}n~Almeida}\ \emph
  {et~al.}(2019)\citenamefont {Beltr{\'a}n~Almeida} \emph
  {et~al.}}]{Almeida:2019iqp}%
  \BibitemOpen
  \bibfield  {author} {\bibinfo {author} {\bibfnamefont {J.~P.}\ \bibnamefont
  {Beltr{\'a}n~Almeida}} \emph {et~al.},\ }\bibfield  {title} {\enquote
  {\bibinfo {title} {{Anisotropic $2$-form dark energy}},}\ }\href {\doibase
  10.1016/j.physletb.2019.05.008} {\bibfield  {journal} {\bibinfo  {journal}
  {Phys. Lett. B}\ }\textbf {\bibinfo {volume} {793}},\ \bibinfo {pages}
  {396--404} (\bibinfo {year} {2019})},\ \Eprint
  {http://arxiv.org/abs/1902.05846} {arXiv:1902.05846 [hep-th]} \BibitemShut
  {NoStop}%
\bibitem [{\citenamefont {Campanelli}\ \emph {et~al.}(2011)\citenamefont
  {Campanelli}, \citenamefont {Cea}, \citenamefont {Fogli},\ and\ \citenamefont
  {Marrone}}]{Campanelli:2010zx}%
  \BibitemOpen
  \bibfield  {author} {\bibinfo {author} {\bibfnamefont {L.}~\bibnamefont
  {Campanelli}}, \bibinfo {author} {\bibfnamefont {P.}~\bibnamefont {Cea}},
  \bibinfo {author} {\bibfnamefont {G.~L.}\ \bibnamefont {Fogli}}, \ and\
  \bibinfo {author} {\bibfnamefont {A.}~\bibnamefont {Marrone}},\ }\bibfield
  {title} {\enquote {\bibinfo {title} {{Testing the Isotropy of the Universe
  with Type Ia Supernovae}},}\ }\href {\doibase 10.1103/PhysRevD.83.103503}
  {\bibfield  {journal} {\bibinfo  {journal} {Phys. Rev. D}\ }\textbf {\bibinfo
  {volume} {83}},\ \bibinfo {pages} {103503} (\bibinfo {year} {2011})},\
  \Eprint {http://arxiv.org/abs/1012.5596} {arXiv:1012.5596 [astro-ph.CO]}
  \BibitemShut {NoStop}%
\bibitem [{\citenamefont {Akarsu}\ \emph {et~al.}(2019)\citenamefont {Akarsu},
  \citenamefont {Kumar}, \citenamefont {Sharma},\ and\ \citenamefont
  {Tedesco}}]{Akarsu:2019pwn}%
  \BibitemOpen
  \bibfield  {author} {\bibinfo {author} {\bibfnamefont {$\ddot{\rm O}$.}\
  \bibnamefont {Akarsu}}, \bibinfo {author} {\bibfnamefont {S.}~\bibnamefont
  {Kumar}}, \bibinfo {author} {\bibfnamefont {S.}~\bibnamefont {Sharma}}, \
  and\ \bibinfo {author} {\bibfnamefont {L.}~\bibnamefont {Tedesco}},\
  }\bibfield  {title} {\enquote {\bibinfo {title} {{Constraints on a Bianchi
  type I spacetime extension of the standard $\Lambda$CDM model}},}\ }\href
  {\doibase 10.1103/PhysRevD.100.023532} {\bibfield  {journal} {\bibinfo
  {journal} {Phys. Rev. D}\ }\textbf {\bibinfo {volume} {100}},\ \bibinfo
  {pages} {023532} (\bibinfo {year} {2019})},\ \Eprint
  {http://arxiv.org/abs/1905.06949} {arXiv:1905.06949 [astro-ph.CO]}
  \BibitemShut {NoStop}%
\bibitem [{\citenamefont {Amirhashchi}\ and\ \citenamefont
  {Amirhashchi}(2020)}]{Amirhashchi:2018nxl}%
  \BibitemOpen
  \bibfield  {author} {\bibinfo {author} {\bibfnamefont {H.}~\bibnamefont
  {Amirhashchi}}\ and\ \bibinfo {author} {\bibfnamefont {S.}~\bibnamefont
  {Amirhashchi}},\ }\bibfield  {title} {\enquote {\bibinfo {title}
  {{Constraining Bianchi Type I Universe With Type Ia Supernova and H(z)
  Data}},}\ }\href {\doibase 10.1016/j.dark.2020.100557} {\bibfield  {journal}
  {\bibinfo  {journal} {Phys. Dark Univ.}\ }\textbf {\bibinfo {volume} {29}},\
  \bibinfo {pages} {100557} (\bibinfo {year} {2020})},\ \Eprint
  {http://arxiv.org/abs/1802.04251} {arXiv:1802.04251 [astro-ph.CO]}
  \BibitemShut {NoStop}%
\bibitem [{\citenamefont {Saadeh}\ \emph {et~al.}(2016)\citenamefont {Saadeh}
  \emph {et~al.}}]{Saadeh:2016sak}%
  \BibitemOpen
  \bibfield  {author} {\bibinfo {author} {\bibfnamefont {D.}~\bibnamefont
  {Saadeh}} \emph {et~al.},\ }\bibfield  {title} {\enquote {\bibinfo {title}
  {{How isotropic is the Universe?}}}\ }\href {\doibase
  10.1103/PhysRevLett.117.131302} {\bibfield  {journal} {\bibinfo  {journal}
  {Phys. Rev. Lett.}\ }\textbf {\bibinfo {volume} {117}},\ \bibinfo {pages}
  {131302} (\bibinfo {year} {2016})},\ \Eprint
  {http://arxiv.org/abs/1605.07178} {arXiv:1605.07178 [astro-ph.CO]}
  \BibitemShut {NoStop}%
\bibitem [{\citenamefont {Amendola}\ \emph {et~al.}(2018)\citenamefont
  {Amendola} \emph {et~al.}}]{Amendola:2016saw}%
  \BibitemOpen
  \bibfield  {author} {\bibinfo {author} {\bibfnamefont {L.}~\bibnamefont
  {Amendola}} \emph {et~al.},\ }\bibfield  {title} {\enquote {\bibinfo {title}
  {{Cosmology and fundamental physics with the Euclid satellite}},}\ }\href
  {\doibase 10.1007/s41114-017-0010-3} {\bibfield  {journal} {\bibinfo
  {journal} {Living Rev. Rel.}\ }\textbf {\bibinfo {volume} {21}},\ \bibinfo
  {pages} {2} (\bibinfo {year} {2018})},\ \Eprint
  {http://arxiv.org/abs/1606.00180} {arXiv:1606.00180 [astro-ph.CO]}
  \BibitemShut {NoStop}%
\bibitem [{\citenamefont {Paliathanasis}\ and\ \citenamefont
  {Leon}(2020)}]{Paliathanasis:2020pax}%
  \BibitemOpen
  \bibfield  {author} {\bibinfo {author} {\bibfnamefont {A.}~\bibnamefont
  {Paliathanasis}}\ and\ \bibinfo {author} {\bibfnamefont {G.}~\bibnamefont
  {Leon}},\ }\bibfield  {title} {\enquote {\bibinfo {title} {{Dynamics and
  exact Bianchi I spacetimes in Einstein\textendash{}\ae{}ther scalar field
  theory}},}\ }\href {\doibase 10.1140/epjc/s10052-020-8148-7} {\bibfield
  {journal} {\bibinfo  {journal} {Eur. Phys. J. C}\ }\textbf {\bibinfo {volume}
  {80}},\ \bibinfo {pages} {589} (\bibinfo {year} {2020})},\ \Eprint
  {http://arxiv.org/abs/2004.08663} {arXiv:2004.08663 [gr-qc]} \BibitemShut
  {NoStop}%
\bibitem [{\citenamefont {Cendra}\ \emph {et~al.}(2001)\citenamefont {Cendra},
  \citenamefont {Marsden},\ and\ \citenamefont {Ratiu}}]{Cendra2001}%
  \BibitemOpen
  \bibfield  {author} {\bibinfo {author} {\bibfnamefont {H.}~\bibnamefont
  {Cendra}}, \bibinfo {author} {\bibfnamefont {J.}~\bibnamefont {Marsden}}, \
  and\ \bibinfo {author} {\bibfnamefont {T.}~\bibnamefont {Ratiu}},\
  }\href@noop {} {\emph {\bibinfo {title} {Geometric Mechanics, Lagrangian
  Reduction, and Nonholonomic Systems}}}\ (\bibinfo  {publisher} {Springer,
  Berlin, Heidelberg},\ \bibinfo {year} {2001})\BibitemShut {NoStop}%
\bibitem [{\citenamefont {Moniz}\ \emph {et~al.}(1993)\citenamefont {Moniz},
  \citenamefont {Mourao},\ and\ \citenamefont {Sa}}]{Moniz:1991kx}%
  \BibitemOpen
  \bibfield  {author} {\bibinfo {author} {\bibfnamefont {P.~V.}\ \bibnamefont
  {Moniz}}, \bibinfo {author} {\bibfnamefont {J.~M.}\ \bibnamefont {Mourao}}, \
  and\ \bibinfo {author} {\bibfnamefont {P.~M.}\ \bibnamefont {Sa}},\
  }\bibfield  {title} {\enquote {\bibinfo {title} {{The Dynamics of a flat
  Friedmann-Robertson-Walker inflationary model in the presence of gauge
  fields}},}\ }\href {\doibase 10.1088/0264-9381/10/3/012} {\bibfield
  {journal} {\bibinfo  {journal} {Class. Quant. Grav.}\ }\textbf {\bibinfo
  {volume} {10}},\ \bibinfo {pages} {517--534} (\bibinfo {year}
  {1993})}\BibitemShut {NoStop}%
\bibitem [{\citenamefont {Ochs}\ and\ \citenamefont
  {Sorg}(1996)}]{Ochs:1996yr}%
  \BibitemOpen
  \bibfield  {author} {\bibinfo {author} {\bibfnamefont {U.}~\bibnamefont
  {Ochs}}\ and\ \bibinfo {author} {\bibfnamefont {M.}~\bibnamefont {Sorg}},\
  }\bibfield  {title} {\enquote {\bibinfo {title} {{Cosmological solutions for
  the coupled Einstein-Yang-Mills-Higgs equations}},}\ }\href {\doibase
  10.1007/BF02107381} {\bibfield  {journal} {\bibinfo  {journal} {Gen. Rel.
  Grav.}\ }\textbf {\bibinfo {volume} {28}},\ \bibinfo {pages} {1177--1219}
  (\bibinfo {year} {1996})}\BibitemShut {NoStop}%
\bibitem [{\citenamefont {Bento}\ \emph {et~al.}(1993)\citenamefont {Bento}
  \emph {et~al.}}]{Bento:1992wy}%
  \BibitemOpen
  \bibfield  {author} {\bibinfo {author} {\bibfnamefont {M.~C.}\ \bibnamefont
  {Bento}} \emph {et~al.},\ }\bibfield  {title} {\enquote {\bibinfo {title}
  {{On the cosmology of massive vector fields with SO(3) global symmetry}},}\
  }\href {\doibase 10.1088/0264-9381/10/2/010} {\bibfield  {journal} {\bibinfo
  {journal} {Class. Quant. Grav.}\ }\textbf {\bibinfo {volume} {10}},\ \bibinfo
  {pages} {285--298} (\bibinfo {year} {1993})},\ \Eprint
  {http://arxiv.org/abs/gr-qc/9302034} {arXiv:gr-qc/9302034 [gr-qc]}
  \BibitemShut {NoStop}%
\bibitem [{\citenamefont {Armendariz-Picon}(2004)}]{ArmendarizPicon:2004pm}%
  \BibitemOpen
  \bibfield  {author} {\bibinfo {author} {\bibfnamefont {C.}~\bibnamefont
  {Armendariz-Picon}},\ }\bibfield  {title} {\enquote {\bibinfo {title} {{Could
  dark energy be vector-like?}}}\ }\href {\doibase
  10.1088/1475-7516/2004/07/007} {\bibfield  {journal} {\bibinfo  {journal}
  {JCAP}\ }\textbf {\bibinfo {volume} {0407}},\ \bibinfo {pages} {007}
  (\bibinfo {year} {2004})},\ \Eprint {http://arxiv.org/abs/astro-ph/0405267}
  {arXiv:astro-ph/0405267} \BibitemShut {NoStop}%
\bibitem [{\citenamefont {Weinberg}(1972)}]{Weinberg:1972kfs}%
  \BibitemOpen
  \bibfield  {author} {\bibinfo {author} {\bibfnamefont {S.}~\bibnamefont
  {Weinberg}},\ }\href@noop {} {\emph {\bibinfo {title} {{Gravitation and
  Cosmology}: {Principles and Applications of the General Theory of
  Relativity}}}}\ (\bibinfo  {publisher} {John Wiley and Sons},\ \bibinfo
  {address} {New York},\ \bibinfo {year} {1972})\BibitemShut {NoStop}%
\bibitem [{\citenamefont {Murata}\ and\ \citenamefont
  {Soda}(2011)}]{Murata:2011wv}%
  \BibitemOpen
  \bibfield  {author} {\bibinfo {author} {\bibfnamefont {K.}~\bibnamefont
  {Murata}}\ and\ \bibinfo {author} {\bibfnamefont {J.}~\bibnamefont {Soda}},\
  }\bibfield  {title} {\enquote {\bibinfo {title} {{Anisotropic Inflation with
  Non-Abelian Gauge Kinetic Function}},}\ }\href {\doibase
  10.1088/1475-7516/2011/06/037} {\bibfield  {journal} {\bibinfo  {journal}
  {JCAP}\ }\textbf {\bibinfo {volume} {1106}},\ \bibinfo {pages} {037}
  (\bibinfo {year} {2011})},\ \Eprint {http://arxiv.org/abs/1103.6164}
  {arXiv:1103.6164 [hep-th]} \BibitemShut {NoStop}%
\bibitem [{\citenamefont {Maleknejad}\ \emph {et~al.}(2012)\citenamefont
  {Maleknejad}, \citenamefont {Sheikh-Jabbari},\ and\ \citenamefont
  {Soda}}]{Maleknejad:2011jr}%
  \BibitemOpen
  \bibfield  {author} {\bibinfo {author} {\bibfnamefont {A.}~\bibnamefont
  {Maleknejad}}, \bibinfo {author} {\bibfnamefont {M.~M.}\ \bibnamefont
  {Sheikh-Jabbari}}, \ and\ \bibinfo {author} {\bibfnamefont {J.}~\bibnamefont
  {Soda}},\ }\bibfield  {title} {\enquote {\bibinfo {title} {{Gauge-flation and
  Cosmic No-Hair Conjecture}},}\ }\href {\doibase
  10.1088/1475-7516/2012/01/016} {\bibfield  {journal} {\bibinfo  {journal}
  {JCAP}\ }\textbf {\bibinfo {volume} {1201}},\ \bibinfo {pages} {016}
  (\bibinfo {year} {2012})},\ \Eprint {http://arxiv.org/abs/1109.5573}
  {arXiv:1109.5573 [hep-th]} \BibitemShut {NoStop}%
\bibitem [{\citenamefont {Maleknejad}\ and\ \citenamefont
  {Erfani}(2014)}]{Maleknejad:2013npa}%
  \BibitemOpen
  \bibfield  {author} {\bibinfo {author} {\bibfnamefont {A.}~\bibnamefont
  {Maleknejad}}\ and\ \bibinfo {author} {\bibfnamefont {E.}~\bibnamefont
  {Erfani}},\ }\bibfield  {title} {\enquote {\bibinfo {title} {{Chromo-Natural
  Model in Anisotropic Background}},}\ }\href {\doibase
  10.1088/1475-7516/2014/03/016} {\bibfield  {journal} {\bibinfo  {journal}
  {JCAP}\ }\textbf {\bibinfo {volume} {1403}},\ \bibinfo {pages} {016}
  (\bibinfo {year} {2014})},\ \Eprint {http://arxiv.org/abs/1311.3361}
  {arXiv:1311.3361 [hep-th]} \BibitemShut {NoStop}%
\bibitem [{\citenamefont {Coley}(2003)}]{Coley:2003mj}%
  \BibitemOpen
  \bibfield  {author} {\bibinfo {author} {\bibfnamefont {A.~A.}\ \bibnamefont
  {Coley}},\ }\href {\doibase 10.1007/978-94-017-0327-7} {\emph {\bibinfo
  {title} {{Dynamical systems and cosmology}}}},\ Vol.\ \bibinfo {volume}
  {291}\ (\bibinfo  {publisher} {Kluwer},\ \bibinfo {address} {Dordrecht,
  Netherlands},\ \bibinfo {year} {2003})\BibitemShut {NoStop}%
\bibitem [{\citenamefont {Wainwright}\ and\ \citenamefont
  {Ellis}(1997)}]{Wainwright2009}%
  \BibitemOpen
  \bibfield  {author} {\bibinfo {author} {\bibfnamefont {J.}~\bibnamefont
  {Wainwright}}\ and\ \bibinfo {author} {\bibfnamefont {G.~F.~R.}\ \bibnamefont
  {Ellis}},\ }\href@noop {} {\emph {\bibinfo {title} {Dynamical Systems in
  Cosmology}}}\ (\bibinfo  {publisher} {Cambridge University Press},\ \bibinfo
  {year} {1997})\BibitemShut {NoStop}%
\bibitem [{\citenamefont {Alho}\ \emph {et~al.}(2020)\citenamefont {Alho},
  \citenamefont {Bessa},\ and\ \citenamefont {Mena}}]{Alho:2019pku}%
  \BibitemOpen
  \bibfield  {author} {\bibinfo {author} {\bibfnamefont {A.}~\bibnamefont
  {Alho}}, \bibinfo {author} {\bibfnamefont {V.}~\bibnamefont {Bessa}}, \ and\
  \bibinfo {author} {\bibfnamefont {F.~C.}\ \bibnamefont {Mena}},\ }\bibfield
  {title} {\enquote {\bibinfo {title} {{Global dynamics of
  Yang\textendash{}Mills field and perfect-fluid Robertson\textendash{}Walker
  cosmologies}},}\ }\href {\doibase 10.1063/1.5139879} {\bibfield  {journal}
  {\bibinfo  {journal} {J. Math. Phys.}\ }\textbf {\bibinfo {volume} {61}},\
  \bibinfo {pages} {032502} (\bibinfo {year} {2020})},\ \Eprint
  {http://arxiv.org/abs/1910.04678} {arXiv:1910.04678 [gr-qc]} \BibitemShut
  {NoStop}%
\bibitem [{\citenamefont {Bean}\ \emph {et~al.}(2001)\citenamefont {Bean},
  \citenamefont {Hansen},\ and\ \citenamefont {Melchiorri}}]{Bean:2001wt}%
  \BibitemOpen
  \bibfield  {author} {\bibinfo {author} {\bibfnamefont {R.}~\bibnamefont
  {Bean}}, \bibinfo {author} {\bibfnamefont {S.~H.}\ \bibnamefont {Hansen}}, \
  and\ \bibinfo {author} {\bibfnamefont {A.}~\bibnamefont {Melchiorri}},\
  }\bibfield  {title} {\enquote {\bibinfo {title} {{Early universe constraints
  on a primordial scaling field}},}\ }\href {\doibase
  10.1103/PhysRevD.64.103508} {\bibfield  {journal} {\bibinfo  {journal} {Phys.
  Rev. D}\ }\textbf {\bibinfo {volume} {64}},\ \bibinfo {pages} {103508}
  (\bibinfo {year} {2001})},\ \Eprint {http://arxiv.org/abs/astro-ph/0104162}
  {arXiv:astro-ph/0104162 [astro-ph]} \BibitemShut {NoStop}%
\bibitem [{\citenamefont {Ade}\ \emph {et~al.}(2016)\citenamefont {Ade} \emph
  {et~al.}}]{Ade:2015rim}%
  \BibitemOpen
  \bibfield  {author} {\bibinfo {author} {\bibfnamefont {P.~A.~R.}\
  \bibnamefont {Ade}} \emph {et~al.} (\bibinfo {collaboration} {Planck}),\
  }\bibfield  {title} {\enquote {\bibinfo {title} {{Planck 2015 results. XIV.
  Dark energy and modified gravity}},}\ }\href {\doibase
  10.1051/0004-6361/201525814} {\bibfield  {journal} {\bibinfo  {journal}
  {Astron. Astrophys.}\ }\textbf {\bibinfo {volume} {594}},\ \bibinfo {pages}
  {A14} (\bibinfo {year} {2016})},\ \Eprint {http://arxiv.org/abs/1502.01590}
  {arXiv:1502.01590 [astro-ph.CO]} \BibitemShut {NoStop}%
\bibitem [{\citenamefont {Zeldovich}(1972)}]{Zeldovich:1972zz}%
  \BibitemOpen
  \bibfield  {author} {\bibinfo {author} {\bibfnamefont {Ya.~B.}\ \bibnamefont
  {Zeldovich}},\ }\bibfield  {title} {\enquote {\bibinfo {title} {{A
  Hypothesis, unifying the structure and the entropy of the universe}},}\
  }\href@noop {} {\bibfield  {journal} {\bibinfo  {journal} {Mon. Not. Roy.
  Astron. Soc.}\ }\textbf {\bibinfo {volume} {160}},\ \bibinfo {pages} {1P--3P}
  (\bibinfo {year} {1972})}\BibitemShut {NoStop}%
\bibitem [{\citenamefont {Mehrabi}\ \emph {et~al.}(2017)\citenamefont
  {Mehrabi}, \citenamefont {Maleknejad},\ and\ \citenamefont
  {Kamali}}]{Mehrabi:2017xga}%
  \BibitemOpen
  \bibfield  {author} {\bibinfo {author} {\bibfnamefont {A.}~\bibnamefont
  {Mehrabi}}, \bibinfo {author} {\bibfnamefont {A.}~\bibnamefont {Maleknejad}},
  \ and\ \bibinfo {author} {\bibfnamefont {V.}~\bibnamefont {Kamali}},\
  }\bibfield  {title} {\enquote {\bibinfo {title} {{Gaugessence: a dark energy
  model with early time radiation-like equation of state}},}\ }\href {\doibase
  10.1007/s10509-017-3033-z} {\bibfield  {journal} {\bibinfo  {journal}
  {Astrophys. Space Sci.}\ }\textbf {\bibinfo {volume} {362}},\ \bibinfo
  {pages} {53} (\bibinfo {year} {2017})},\ \Eprint
  {http://arxiv.org/abs/1510.00838} {arXiv:1510.00838 [astro-ph.CO]}
  \BibitemShut {NoStop}%
\bibitem [{\citenamefont {Huterer}\ \emph {et~al.}(2015)\citenamefont {Huterer}
  \emph {et~al.}}]{Huterer:2013xky}%
  \BibitemOpen
  \bibfield  {author} {\bibinfo {author} {\bibfnamefont {D.}~\bibnamefont
  {Huterer}} \emph {et~al.},\ }\bibfield  {title} {\enquote {\bibinfo {title}
  {{Growth of Cosmic Structure: Probing Dark Energy Beyond Expansion}},}\
  }\href {\doibase 10.1016/j.astropartphys.2014.07.004} {\bibfield  {journal}
  {\bibinfo  {journal} {Astropart. Phys.}\ }\textbf {\bibinfo {volume} {63}},\
  \bibinfo {pages} {23--41} (\bibinfo {year} {2015})},\ \Eprint
  {http://arxiv.org/abs/1309.5385} {arXiv:1309.5385 [astro-ph.CO]} \BibitemShut
  {NoStop}%
\bibitem [{\citenamefont {Golovnev}\ \emph {et~al.}(2008)\citenamefont
  {Golovnev}, \citenamefont {Mukhanov},\ and\ \citenamefont
  {Vanchurin}}]{Golovnev:2008cf}%
  \BibitemOpen
  \bibfield  {author} {\bibinfo {author} {\bibfnamefont {A.}~\bibnamefont
  {Golovnev}}, \bibinfo {author} {\bibfnamefont {V.}~\bibnamefont {Mukhanov}},
  \ and\ \bibinfo {author} {\bibfnamefont {V.}~\bibnamefont {Vanchurin}},\
  }\bibfield  {title} {\enquote {\bibinfo {title} {{Vector Inflation}},}\
  }\href {\doibase 10.1088/1475-7516/2008/06/009} {\bibfield  {journal}
  {\bibinfo  {journal} {JCAP}\ }\textbf {\bibinfo {volume} {0806}},\ \bibinfo
  {pages} {009} (\bibinfo {year} {2008})},\ \Eprint
  {http://arxiv.org/abs/0802.2068} {arXiv:0802.2068 [astro-ph]} \BibitemShut
  {NoStop}%
\bibitem [{\citenamefont {Maleknejad}\ and\ \citenamefont
  {Sheikh-Jabbari}(2013)}]{Maleknejad:2011jw}%
  \BibitemOpen
  \bibfield  {author} {\bibinfo {author} {\bibfnamefont {A.}~\bibnamefont
  {Maleknejad}}\ and\ \bibinfo {author} {\bibfnamefont {M.~M.}\ \bibnamefont
  {Sheikh-Jabbari}},\ }\bibfield  {title} {\enquote {\bibinfo {title}
  {{Gauge-flation: Inflation From Non-Abelian Gauge Fields}},}\ }\href
  {\doibase 10.1016/j.physletb.2013.05.001} {\bibfield  {journal} {\bibinfo
  {journal} {Phys. Lett. B}\ }\textbf {\bibinfo {volume} {723}},\ \bibinfo
  {pages} {224--228} (\bibinfo {year} {2013})},\ \Eprint
  {http://arxiv.org/abs/1102.1513} {arXiv:1102.1513 [hep-ph]} \BibitemShut
  {NoStop}%
\bibitem [{\citenamefont {Adshead}\ and\ \citenamefont
  {Wyman}(2012)}]{Adshead:2012kp}%
  \BibitemOpen
  \bibfield  {author} {\bibinfo {author} {\bibfnamefont {P.}~\bibnamefont
  {Adshead}}\ and\ \bibinfo {author} {\bibfnamefont {M.}~\bibnamefont
  {Wyman}},\ }\bibfield  {title} {\enquote {\bibinfo {title} {{Chromo-Natural
  Inflation: Natural inflation on a steep potential with classical non-Abelian
  gauge fields}},}\ }\href {\doibase 10.1103/PhysRevLett.108.261302} {\bibfield
   {journal} {\bibinfo  {journal} {Phys. Rev. Lett.}\ }\textbf {\bibinfo
  {volume} {108}},\ \bibinfo {pages} {261302} (\bibinfo {year} {2012})},\
  \Eprint {http://arxiv.org/abs/1202.2366} {arXiv:1202.2366 [hep-th]}
  \BibitemShut {NoStop}%
\bibitem [{\citenamefont {Rodriguez}\ and\ \citenamefont
  {Navarro}(2018)}]{Rodriguez:2017wkg}%
  \BibitemOpen
  \bibfield  {author} {\bibinfo {author} {\bibfnamefont {Y.}~\bibnamefont
  {Rodriguez}}\ and\ \bibinfo {author} {\bibfnamefont {A.~A.}\ \bibnamefont
  {Navarro}},\ }\bibfield  {title} {\enquote {\bibinfo {title} {{Non-Abelian
  $S$-term dark energy and inflation}},}\ }\href {\doibase
  10.1016/j.dark.2018.01.003} {\bibfield  {journal} {\bibinfo  {journal} {Phys.
  Dark Univ.}\ }\textbf {\bibinfo {volume} {19}},\ \bibinfo {pages} {129--136}
  (\bibinfo {year} {2018})},\ \Eprint {http://arxiv.org/abs/1711.01935}
  {arXiv:1711.01935 [gr-qc]} \BibitemShut {NoStop}%
\bibitem [{\citenamefont {Gomez}\ and\ \citenamefont
  {Rodriguez}(2020)}]{Gomez:2020sfz}%
  \BibitemOpen
  \bibfield  {author} {\bibinfo {author} {\bibfnamefont {L.~G.}\ \bibnamefont
  {Gomez}}\ and\ \bibinfo {author} {\bibfnamefont {Y.}~\bibnamefont
  {Rodriguez}},\ }\bibfield  {title} {\enquote {\bibinfo {title} {{Coupled
  Multi-Proca Vector Dark Energy}},}\ }\href@noop {} {\  (\bibinfo {year}
  {2020})},\ \Eprint {http://arxiv.org/abs/2004.06466} {arXiv:2004.06466
  [gr-qc]} \BibitemShut {NoStop}%
\bibitem [{\citenamefont {Endlich}\ \emph {et~al.}(2013)\citenamefont
  {Endlich}, \citenamefont {Nicolis},\ and\ \citenamefont
  {Wang}}]{Endlich:2012pz}%
  \BibitemOpen
  \bibfield  {author} {\bibinfo {author} {\bibfnamefont {S.}~\bibnamefont
  {Endlich}}, \bibinfo {author} {\bibfnamefont {A.}~\bibnamefont {Nicolis}}, \
  and\ \bibinfo {author} {\bibfnamefont {J.}~\bibnamefont {Wang}},\ }\bibfield
  {title} {\enquote {\bibinfo {title} {{Solid Inflation}},}\ }\href {\doibase
  10.1088/1475-7516/2013/10/011} {\bibfield  {journal} {\bibinfo  {journal}
  {JCAP}\ }\textbf {\bibinfo {volume} {1310}},\ \bibinfo {pages} {011}
  (\bibinfo {year} {2013})},\ \Eprint {http://arxiv.org/abs/1210.0569}
  {arXiv:1210.0569 [hep-th]} \BibitemShut {NoStop}%
\bibitem [{\citenamefont {Armendariz-Picon}(2007)}]{ArmendarizPicon:2007nr}%
  \BibitemOpen
  \bibfield  {author} {\bibinfo {author} {\bibfnamefont {C.}~\bibnamefont
  {Armendariz-Picon}},\ }\bibfield  {title} {\enquote {\bibinfo {title}
  {{Creating Statistically Anisotropic and Inhomogeneous Perturbations}},}\
  }\href {\doibase 10.1088/1475-7516/2007/09/014} {\bibfield  {journal}
  {\bibinfo  {journal} {JCAP}\ }\textbf {\bibinfo {volume} {0709}},\ \bibinfo
  {pages} {014} (\bibinfo {year} {2007})},\ \Eprint
  {http://arxiv.org/abs/0705.1167} {arXiv:0705.1167 [astro-ph]} \BibitemShut
  {NoStop}%
\bibitem [{\citenamefont {Firouzjahi}\ \emph {et~al.}(2019)\citenamefont
  {Firouzjahi} \emph {et~al.}}]{Firouzjahi:2018wlp}%
  \BibitemOpen
  \bibfield  {author} {\bibinfo {author} {\bibfnamefont {H.}~\bibnamefont
  {Firouzjahi}} \emph {et~al.},\ }\bibfield  {title} {\enquote {\bibinfo
  {title} {{Charged Vector Inflation}},}\ }\href {\doibase
  10.1103/PhysRevD.100.043530} {\bibfield  {journal} {\bibinfo  {journal}
  {Phys.\ Rev.\ D}\ }\textbf {\bibinfo {volume} {100}},\ \bibinfo {pages}
  {043530} (\bibinfo {year} {2019})},\ \Eprint
  {http://arxiv.org/abs/1812.07464} {arXiv:1812.07464 [hep-th]} \BibitemShut
  {NoStop}%
\bibitem [{\citenamefont {Amendola}\ and\ \citenamefont
  {Tsujikawa}(2015)}]{Amendola:2015ksp}%
  \BibitemOpen
  \bibfield  {author} {\bibinfo {author} {\bibfnamefont {L.}~\bibnamefont
  {Amendola}}\ and\ \bibinfo {author} {\bibfnamefont {S.}~\bibnamefont
  {Tsujikawa}},\ }\href
  {http://www.cambridge.org/academic/subjects/physics/cosmology-relativity-and-gravitation/dark-energy-theory-and-observations?format=PB&isbn=9781107453982}
  {\emph {\bibinfo {title} {{Dark Energy}}}}\ (\bibinfo  {publisher} {Cambridge
  University Press},\ \bibinfo {year} {2015})\BibitemShut {NoStop}%
\bibitem [{\citenamefont {Ferreira}\ and\ \citenamefont
  {Joyce}(1998)}]{Ferreira:1997hj}%
  \BibitemOpen
  \bibfield  {author} {\bibinfo {author} {\bibfnamefont {P.~G.}\ \bibnamefont
  {Ferreira}}\ and\ \bibinfo {author} {\bibfnamefont {M.}~\bibnamefont
  {Joyce}},\ }\bibfield  {title} {\enquote {\bibinfo {title} {{Cosmology with a
  primordial scaling field}},}\ }\href {\doibase 10.1103/PhysRevD.58.023503}
  {\bibfield  {journal} {\bibinfo  {journal} {Phys. Rev. D}\ }\textbf {\bibinfo
  {volume} {58}},\ \bibinfo {pages} {023503} (\bibinfo {year} {1998})},\
  \Eprint {http://arxiv.org/abs/astro-ph/9711102} {arXiv:astro-ph/9711102}
  \BibitemShut {NoStop}%
\bibitem [{\citenamefont {Pallis}(2006)}]{Pallis:2005bb}%
  \BibitemOpen
  \bibfield  {author} {\bibinfo {author} {\bibfnamefont {C.}~\bibnamefont
  {Pallis}},\ }\bibfield  {title} {\enquote {\bibinfo {title}
  {{Kination-dominated reheating and cold dark matter abundance}},}\ }\href
  {\doibase 10.1016/j.nuclphysb.2006.06.003} {\bibfield  {journal} {\bibinfo
  {journal} {Nucl. Phys. B}\ }\textbf {\bibinfo {volume} {751}},\ \bibinfo
  {pages} {129--159} (\bibinfo {year} {2006})},\ \Eprint
  {http://arxiv.org/abs/hep-ph/0510234} {arXiv:hep-ph/0510234} \BibitemShut
  {NoStop}%
\bibitem [{\citenamefont {Dimopoulos}\ and\ \citenamefont
  {Markkanen}(2018)}]{Dimopoulos:2018wfg}%
  \BibitemOpen
  \bibfield  {author} {\bibinfo {author} {\bibfnamefont {K.}~\bibnamefont
  {Dimopoulos}}\ and\ \bibinfo {author} {\bibfnamefont {T.}~\bibnamefont
  {Markkanen}},\ }\bibfield  {title} {\enquote {\bibinfo {title} {{Non-minimal
  gravitational reheating during kination}},}\ }\href {\doibase
  10.1088/1475-7516/2018/06/021} {\bibfield  {journal} {\bibinfo  {journal}
  {JCAP}\ }\textbf {\bibinfo {volume} {1806}},\ \bibinfo {pages} {021}
  (\bibinfo {year} {2018})},\ \Eprint {http://arxiv.org/abs/1803.07399}
  {arXiv:1803.07399 [gr-qc]} \BibitemShut {NoStop}%
\end{thebibliography}%

\end{document}